\documentclass[final]{IEEEtran}

\usepackage{graphicx}
\usepackage{subfigure}
\usepackage{multirow}
\usepackage{array}
\newcolumntype{P}[1]{>{\centering\arraybackslash}p{#1}}
\newcolumntype{M}[1]{>{\centering\arraybackslash}m{#1}}

\usepackage{amsthm,amssymb,amsmath,graphicx,multirow,color,amsfonts}%
\usepackage[update,prepend]{epstopdf}
\usepackage{multirow}
\usepackage[latin1]{inputenc}
\usepackage{tikz}
\usepackage{bbm} % for \mathbbm{1}
\usepackage{pdfpages}
\usepackage{multirow}
\usepackage{subfig}
\usepackage{comment}
 \usepackage{graphicx}
\usepackage{multicol}
\usepackage{cite}

\usepackage[justification=centering]{caption}
\usepackage{textcomp}
\usepackage{psfrag}
\usepackage{arydshln}
\usepackage{url}
\usepackage{soul}
\usepackage{graphicx,color}
\usepackage[nolist]{acronym}
\usepackage{algorithm,algorithmic} %algorithm
\usepackage{mathtools,lipsum}
\usepackage{cuted}
\usepackage{amsmath}
\newtheorem{proposition}{Proposition} 
\usepackage{mathrsfs}

\usepackage[capitalise]{cleveref}
\Crefname{equation}{Eq.\!}{Eqs.\!}
\Crefname{figure}{Fig.\!}{Figs.\!}
\Crefname{tabular}{Tab.\!}{Tabs.\!}
\Crefname{section}{Section\!}{Sections.\!}

\def\nb0{{\mathbf{0}}}
\def\nb1{{\mathbf{1}}}

\newtheorem{lemma}{Lemma}

\newtheorem{definition}{Definition}

\newtheorem{theorem}{Theorem}

\newtheorem{remark}{Remark}

\newenvironment{sequation}{
\begin{equation}\small}{\end{equation}
}

\begin{document}
%\pagenumbering{gobble}
\graphicspath{{./Figures/}}
	\begin{acronym}

\acro{5G-NR}{5G New Radio}
\acro{3GPP}{3rd Generation Partnership Project}
\acro{ABS}{aerial base station}
\acro{AC}{address coding}
\acro{ACF}{autocorrelation function}
\acro{ACR}{autocorrelation receiver}
\acro{ADC}{analog-to-digital converter}
\acrodef{aic}[AIC]{Analog-to-Information Converter}     
\acro{AIC}[AIC]{Akaike information criterion}
\acro{aric}[ARIC]{asymmetric restricted isometry constant}
\acro{arip}[ARIP]{asymmetric restricted isometry property}

\acro{ARQ}{Automatic Repeat Request}
\acro{AUB}{asymptotic union bound}
\acrodef{awgn}[AWGN]{Additive White Gaussian Noise}     
\acro{AWGN}{additive white Gaussian noise}

\acro{APSK}[PSK]{asymmetric PSK} 

\acro{waric}[AWRICs]{asymmetric weak restricted isometry constants}
\acro{warip}[AWRIP]{asymmetric weak restricted isometry property}
\acro{BCH}{Bose, Chaudhuri, and Hocquenghem}        
\acro{BCHC}[BCHSC]{BCH based source coding}
\acro{BEP}{bit error probability}
\acro{BFC}{block fading channel}
\acro{BG}[BG]{Bernoulli-Gaussian}
\acro{BGG}{Bernoulli-Generalized Gaussian}
\acro{BPAM}{binary pulse amplitude modulation}
\acro{BPDN}{Basis Pursuit Denoising}
\acro{BPPM}{binary pulse position modulation}
\acro{BPSK}{Binary Phase Shift Keying}
\acro{BPZF}{bandpass zonal filter}
\acro{BSC}{binary symmetric channels}              
\acro{BU}[BU]{Bernoulli-uniform}
\acro{BER}{bit error rate}
\acro{BS}{base station}
\acro{BW}{BandWidth}
\acro{BLLL}{ binary log-linear learning }

\acro{CP}{Cyclic Prefix}
\acrodef{cdf}[CDF]{cumulative distribution function}   
\acro{CDF}{Cumulative Distribution Function}
\acrodef{c.d.f.}[CDF]{cumulative distribution function}
\acro{CCDF}{complementary cumulative distribution function}
\acrodef{ccdf}[CCDF]{complementary CDF}               
\acrodef{c.c.d.f.}[CCDF]{complementary cumulative distribution function}
\acro{CD}{cooperative diversity}

\acro{CDMA}{Code Division Multiple Access}
\acro{ch.f.}{characteristic function}
\acro{CIR}{channel impulse response}
\acro{cosamp}[CoSaMP]{compressive sampling matching pursuit}
\acro{CR}{cognitive radio}
\acro{cs}[CS]{compressed sensing}                   
\acrodef{cscapital}[CS]{Compressed sensing} %will not include it in the list
\acrodef{CS}[CS]{compressed sensing}
\acro{CSI}{channel state information}
\acro{CCSDS}{consultative committee for space data systems}
\acro{CC}{convolutional coding}
\acro{Covid19}[COVID-19]{Coronavirus disease}

\acro{DAA}{detect and avoid}
\acro{DAB}{digital audio broadcasting}
\acro{DCT}{discrete cosine transform}
\acro{dft}[DFT]{discrete Fourier transform}
\acro{DR}{distortion-rate}
\acro{DS}{direct sequence}
\acro{DS-SS}{direct-sequence spread-spectrum}
\acro{DTR}{differential transmitted-reference}
\acro{DVB-H}{digital video broadcasting\,--\,handheld}
\acro{DVB-T}{digital video broadcasting\,--\,terrestrial}
\acro{DL}{DownLink}
\acro{DSSS}{Direct Sequence Spread Spectrum}
\acro{DFT-s-OFDM}{Discrete Fourier Transform-spread-Orthogonal Frequency Division Multiplexing}
\acro{DAS}{Distributed Antenna System}
\acro{DNA}{DeoxyriboNucleic Acid}

\acro{EC}{European Commission}
\acro{EED}[EED]{exact eigenvalues distribution}
\acro{EIRP}{Equivalent Isotropically Radiated Power}
\acro{ELP}{equivalent low-pass}
\acro{eMBB}{Enhanced Mobile Broadband}
\acro{EMF}{ElectroMagnetic Field}
\acro{EU}{European union}
\acro{EI}{Exposure Index}
\acro{eICIC}{enhanced Inter-Cell Interference Coordination}

\acro{FC}[FC]{fusion center}
\acro{FCC}{Federal Communications Commission}
\acro{FEC}{forward error correction}
\acro{FFT}{fast Fourier transform}
\acro{FH}{frequency-hopping}
\acro{FH-SS}{frequency-hopping spread-spectrum}
\acrodef{FS}{Frame synchronization}
\acro{FSsmall}[FS]{frame synchronization}  
\acro{FDMA}{Frequency Division Multiple Access}

\acro{GA}{Gaussian approximation}
\acro{GF}{Galois field }
\acro{GG}{Generalized-Gaussian}
\acro{GIC}[GIC]{generalized information criterion}
\acro{GLRT}{generalized likelihood ratio test}
\acro{GPS}{Global Positioning System}
\acro{GMSK}{Gaussian Minimum Shift Keying}
\acro{GSMA}{Global System for Mobile communications Association}
\acro{GS}{ground station}
\acro{GMG}{ Grid-connected MicroGeneration}

\acro{HAP}{high altitude platform}
\acro{HetNet}{Heterogeneous network}

\acro{IDR}{information distortion-rate}
\acro{IFFT}{inverse fast Fourier transform}
\acro{iht}[IHT]{iterative hard thresholding}
\acro{i.i.d.}{independent, identically distributed}
\acro{IoT}{Internet of Things}                      
\acro{IR}{impulse radio}
\acro{lric}[LRIC]{lower restricted isometry constant}
\acro{lrict}[LRICt]{lower restricted isometry constant threshold}
\acro{ISI}{intersymbol interference}
\acro{ITU}{International Telecommunication Union}
\acro{ICNIRP}{International Commission on Non-Ionizing Radiation Protection}
\acro{IEEE}{Institute of Electrical and Electronics Engineers}
\acro{ICES}{IEEE international committee on electromagnetic safety}
\acro{IEC}{International Electrotechnical Commission}
\acro{IARC}{International Agency on Research on Cancer}
\acro{IS-95}{Interim Standard 95}

\acro{KPI}{Key Performance Indicator}

\acro{LEO}{low earth orbit}
\acro{LF}{likelihood function}
\acro{LLF}{log-likelihood function}
\acro{LLR}{log-likelihood ratio}
\acro{LLRT}{log-likelihood ratio test}
\acro{LoS}{Line-of-Sight}
\acro{LRT}{likelihood ratio test}
\acro{wlric}[LWRIC]{lower weak restricted isometry constant}
\acro{wlrict}[LWRICt]{LWRIC threshold}
\acro{LPWAN}{Low Power Wide Area Network}
\acro{LoRaWAN}{Low power long Range Wide Area Network}
\acro{NLoS}{Non-Line-of-Sight}
\acro{LiFi}[Li-Fi]{light-fidelity}
 \acro{LED}{light emitting diode}
 \acro{LABS}{LoS transmission with each ABS}
 \acro{NLABS}{NLoS transmission with each ABS}

\acro{MB}{multiband}
\acro{MC}{macro cell}
\acro{MDS}{mixed distributed source}
\acro{MF}{matched filter}
\acro{m.g.f.}{moment generating function}
\acro{MI}{mutual information}
\acro{MIMO}{Multiple-Input Multiple-Output}
\acro{MISO}{multiple-input single-output}
\acrodef{maxs}[MJSO]{maximum joint support cardinality}                       
\acro{ML}[ML]{maximum likelihood}
\acro{MMSE}{minimum mean-square error}
\acro{MMV}{multiple measurement vectors}
\acrodef{MOS}{model order selection}
\acro{M-PSK}[${M}$-PSK]{$M$-ary phase shift keying}                       
\acro{M-APSK}[${M}$-PSK]{$M$-ary asymmetric PSK} 
\acro{MP}{ multi-period}
\acro{MINLP}{mixed integer non-linear programming}

\acro{M-QAM}[$M$-QAM]{$M$-ary quadrature amplitude modulation}
\acro{MRC}{maximal ratio combiner}                  
\acro{maxs}[MSO]{maximum sparsity order}                                      
\acro{M2M}{Machine-to-Machine}                                                
\acro{MUI}{multi-user interference}
\acro{mMTC}{massive Machine Type Communications}      
\acro{mm-Wave}{millimeter-wave}
\acro{MP}{mobile phone}
\acro{MPE}{maximum permissible exposure}
\acro{MAC}{media access control}
\acro{NB}{narrowband}
\acro{NBI}{narrowband interference}
\acro{NLA}{nonlinear sparse approximation}
\acro{NLOS}{Non-Line of Sight}
\acro{NTIA}{National Telecommunications and Information Administration}
\acro{NTP}{National Toxicology Program}
\acro{NHS}{National Health Service}

\acro{LOS}{Line of Sight}

\acro{OC}{optimum combining}                             
\acro{OC}{optimum combining}
\acro{ODE}{operational distortion-energy}
\acro{ODR}{operational distortion-rate}
\acro{OFDM}{Orthogonal Frequency-Division Multiplexing}
\acro{omp}[OMP]{orthogonal matching pursuit}
\acro{OSMP}[OSMP]{orthogonal subspace matching pursuit}
\acro{OQAM}{offset quadrature amplitude modulation}
\acro{OQPSK}{offset QPSK}
\acro{OFDMA}{Orthogonal Frequency-division Multiple Access}
\acro{OPEX}{Operating Expenditures}
\acro{OQPSK/PM}{OQPSK with phase modulation}

\acro{PAM}{pulse amplitude modulation}
\acro{PAR}{peak-to-average ratio}
\acrodef{pdf}[PDF]{probability density function}                      
\acro{PDF}{probability density function}
\acrodef{p.d.f.}[PDF]{probability distribution function}
\acro{PDP}{power dispersion profile}
\acro{PMF}{probability mass function}                             
\acrodef{p.m.f.}[PMF]{probability mass function}
\acro{PN}{pseudo-noise}
\acro{PPM}{pulse position modulation}
\acro{PRake}{Partial Rake}
\acro{PSD}{power spectral density}
\acro{PSEP}{pairwise synchronization error probability}
\acro{PSK}{phase shift keying}
\acro{PD}{power density}
\acro{8-PSK}[$8$-PSK]{$8$-phase shift keying}
\acro{PPP}{Poisson point process}
\acro{PCP}{Poisson cluster process}
 
\acro{FSK}{Frequency Shift Keying}

\acro{QAM}{Quadrature Amplitude Modulation}
\acro{QPSK}{Quadrature Phase Shift Keying}
\acro{OQPSK/PM}{OQPSK with phase modulator }

\acro{RD}[RD]{raw data}
\acro{RDL}{"random data limit"}
\acro{ric}[RIC]{restricted isometry constant}
\acro{rict}[RICt]{restricted isometry constant threshold}
\acro{rip}[RIP]{restricted isometry property}
\acro{ROC}{receiver operating characteristic}
\acro{rq}[RQ]{Raleigh quotient}
\acro{RS}[RS]{Reed-Solomon}
\acro{RSC}[RSSC]{RS based source coding}
\acro{r.v.}{random variable}                               
\acro{R.V.}{random vector}
\acro{RMS}{root mean square}
\acro{RFR}{radiofrequency radiation}
\acro{RIS}{Reconfigurable Intelligent Surface}
\acro{RNA}{RiboNucleic Acid}
\acro{RRM}{Radio Resource Management}
\acro{RUE}{reference user equipments}
\acro{RAT}{radio access technology}
\acro{RB}{resource block}

\acro{SA}[SA-Music]{subspace-augmented MUSIC with OSMP}
\acro{SC}{small cell}
\acro{SCBSES}[SCBSES]{Source Compression Based Syndrome Encoding Scheme}
\acro{SCM}{sample covariance matrix}
\acro{SEP}{symbol error probability}
\acro{SG}[SG]{sparse-land Gaussian model}
\acro{SIMO}{single-input multiple-output}
\acro{SINR}{signal-to-interference plus noise ratio}
\acro{SIR}{signal-to-interference ratio}
\acro{SISO}{Single-Input Single-Output}
\acro{SMV}{single measurement vector}
\acro{SNR}[\textrm{SNR}]{signal-to-noise ratio} 
\acro{sp}[SP]{subspace pursuit}
\acro{SS}{spread spectrum}
\acro{SW}{sync word}
\acro{SAR}{specific absorption rate}
\acro{SSB}{synchronization signal block}
\acro{SR}{shrink and realign}

\acro{tUAV}{tethered Unmanned Aerial Vehicle}
\acro{TBS}{terrestrial base station}

\acro{uUAV}{untethered Unmanned Aerial Vehicle}
\acro{PDF}{probability density functions}

\acro{PL}{path-loss}

\acro{TH}{time-hopping}
\acro{ToA}{time-of-arrival}
\acro{TR}{transmitted-reference}
\acro{TW}{Tracy-Widom}
\acro{TWDT}{TW Distribution Tail}
\acro{TCM}{trellis coded modulation}
\acro{TDD}{Time-Division Duplexing}
\acro{TDMA}{Time Division Multiple Access}
\acro{Tx}{average transmit}

\acro{UAV}{Unmanned Aerial Vehicle}
\acro{uric}[URIC]{upper restricted isometry constant}
\acro{urict}[URICt]{upper restricted isometry constant threshold}
\acro{UWB}{ultrawide band}
\acro{UWBcap}[UWB]{Ultrawide band}   
\acro{URLLC}{Ultra Reliable Low Latency Communications}
         
\acro{wuric}[UWRIC]{upper weak restricted isometry constant}
\acro{wurict}[UWRICt]{UWRIC threshold}                
\acro{UE}{User Equipment}
\acro{UL}{UpLink}

\acro{WiM}[WiM]{weigh-in-motion}
\acro{WLAN}{wireless local area network}
\acro{wm}[WM]{Wishart matrix}                               
\acroplural{wm}[WM]{Wishart matrices}
\acro{WMAN}{wireless metropolitan area network}
\acro{WPAN}{wireless personal area network}
\acro{wric}[WRIC]{weak restricted isometry constant}
\acro{wrict}[WRICt]{weak restricted isometry constant thresholds}
\acro{wrip}[WRIP]{weak restricted isometry property}
\acro{WSN}{wireless sensor network}                        
\acro{WSS}{Wide-Sense Stationary}
\acro{WHO}{World Health Organization}
\acro{Wi-Fi}{Wireless Fidelity}

\acro{sss}[SpaSoSEnc]{sparse source syndrome encoding}

\acro{VLC}{Visible Light Communication}
\acro{VPN}{Virtual Private Network} 
\acro{RF}{Radio Frequency}
\acro{FSO}{Free Space Optics}
\acro{IoST}{Internet of Space Things}

\acro{GSM}{Global System for Mobile Communications}
\acro{2G}{Second-generation cellular network}
\acro{3G}{Third-generation cellular network}
\acro{4G}{Fourth-generation cellular network}
\acro{5G}{Fifth-generation cellular network}	
\acro{gNB}{next-generation Node-B Base Station}
\acro{NR}{New Radio}
\acro{UMTS}{Universal Mobile Telecommunications Service}
\acro{LTE}{Long Term Evolution}

\acro{QoS}{Quality of Service}
\end{acronym}
	
	%% EMF definitions
\newcommand{\SAR} {\mathrm{SAR}}
\newcommand{\WBSAR} {\mathrm{SAR}_{\mathsf{WB}}}
\newcommand{\gSAR} {\mathrm{SAR}_{10\si{\gram}}}
\newcommand{\Sab} {S_{\mathsf{ab}}}
\newcommand{\Eavg} {E_{\mathsf{avg}}}
\newcommand{\ft}{f_{\textsf{th}}}
\newcommand{\alphatf}{\alpha_{24}}

\title{
Satellite-Terrestrial Routing or Inter-Satellite Routing? A Stochastic Geometry Perspective
}

\author{
Ruibo Wang, {\em Member, IEEE}, Mustafa A. Kishk, {\em Member, IEEE},\\ 
Howard H. Yang, {\em Member, IEEE}, and Mohamed-Slim Alouini, {\em Fellow, IEEE}
\thanks{Ruibo Wang and Mohamed-Slim Alouini are with King Abdullah University of Science and Technology (KAUST), CEMSE division, Thuwal 23955-6900, Saudi Arabia. Mustafa A. Kishk is with the Department of Electronic Engineering, Maynooth University, Maynooth, W23 F2H6, Ireland. Howard H. Yang is with the ZJU-UIUC Institute, Zhejiang University, Haining 314400, China. (e-mail: ruibo.wang@kaust.edu.sa; mustafa.kishk@mu.ie; haoyang@intl.zju.edu.cn; slim.alouini@kaust.edu.sa). Corresponding author: Howard H. Yang.
}
\vspace{-8mm}
}
\maketitle

{\color{black}
\vspace{-4mm}
\begin{abstract}
The design and comparison of satellite-terrestrial routing (STR) and inter-satellite routing (ISR) in low Earth orbit satellite constellations is a widely discussed topic. The signal propagation distance under STR is generally longer than that under ISR, resulting in greater path loss. The global deployment of gateways introduces additional costs for STR. In contrast, transmissions under ISR rely on the energy of satellites, which could be more costly. Additionally, ISLs require more complex communication protocol design, extra hardware support, and increased computational power.
To maximize energy efficiency, we propose two optimal routing relay selection algorithms for ISR and STR, respectively. Furthermore, we derive the analytical expressions for the routing availability probability and energy efficiency, quantifying the performance of the algorithms. 
The analyses enable us to assess the performance of the proposed algorithms against existing methods through numerical results, compare the performance of STR and ISR, and provide useful insights for constellation design. 
% In numerical results, we assess the performance of the proposed algorithms against existing methods, compare the performance of STR and ISR, and provide insights for constellation design. 
\end{abstract}
}

\begin{IEEEkeywords}
Inter-satellite routing, satellite-terrestrial routing, energy efficiency, stochastic geometry.
\end{IEEEkeywords}

\section{Introduction}
\begin{figure}[ht]
\centering
\includegraphics[width=0.9\linewidth]{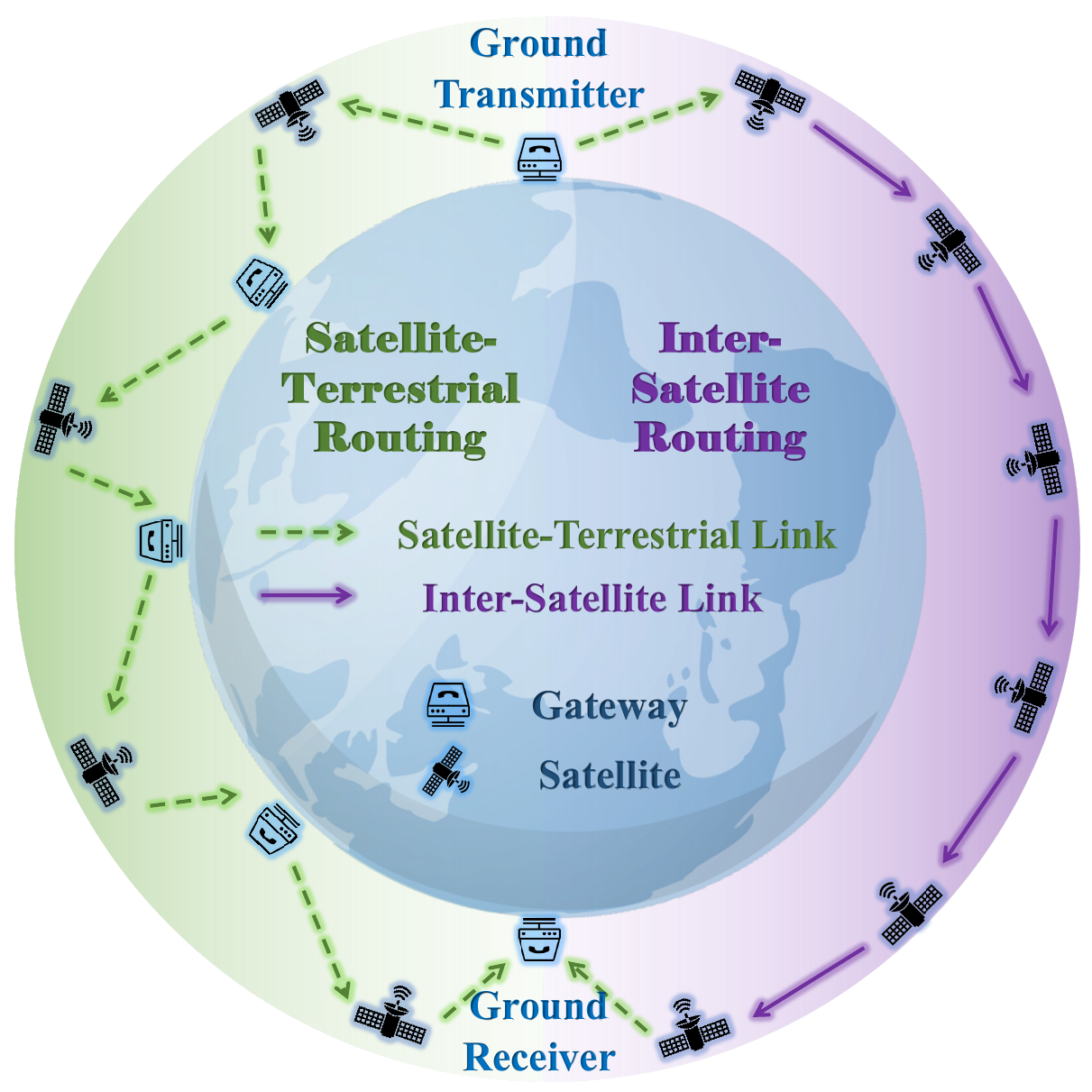}
\caption{An illustrative comparison between the satellite-terrestrial routing and inter-satellite routing approaches.}
\label{figure1}
\end{figure}

In recent years, we have witnessed an explosion in the development of low Earth orbit (LEO) satellite constellations \cite{yue2022security}. 
Due to LEO satellites' salient advantages in ultra-long distance communications \cite{chaudhry2020free}, the LEO satellite network is expected to be pivotal in the next-generation wireless network. As shown in Fig.~\ref{figure1}, the long-distance satellite routing can further be divided into satellite-terrestrial routing (STR) and inter-satellite routing (ISR) based on the selection of a ground gateway (GW) or another satellite as the next relay by satellites \cite{zhu2021integrated}. Compared to satellite-terrestrial links (STLs), inter-satellite links (ISLs) are less affected by shadowing and multipath effects, and applying ISLs can reduce reliance on ground-based infrastructure. Hence, many constellations, including Starlink, Kuiper, and Telesat, claim to support ISLs \cite{chaudhry2022temporary}. However, ISLs require more complex communication protocol design, additional hardware support, and increased computational power and energy demands on the satellite \cite{radhakrishnan2016survey}. Constellations like OneWeb and Globalstar do not configure ISLs \cite{del2019technical}. As a result, whether to rely more on ISLs and fully entrust a greater portion of communication tasks to space is a highly debated issue.

\par
% Comparing STR and ISR from the perspective of routing performance is intriguing. 
A comprehensive comparison between STR and ISR from the perspective of routing performance is crucial for network designs. 
Specifically, STR's bent-pipe satellite-terrestrial round trip inevitably extends the communication distance, leading to greater path loss \cite{wang2022ultra}, requiring more energy consumption. 
In contrast, ISR consumes less energy for the same transmission task but relies more on the satellite's precious energy resources. In this article, we aim to quantitatively compare the communication costs of STR and ISR. Further details about the challenges faced in quantitative comparison and how to compare the communication costs are provided in Sec.~\ref{secI-A} and Sec.~\ref{secI-B}, respectively.

\subsection{Related Works}\label{secI-A}
The majority of existing routing methods are not applicable for quantitatively evaluating the performance of STR and ISR in LEO satellite constellations. They are mainly divided into graph theory-based routing analysis \cite{9348676_1,shen2020dynamic_2,geng2021agent_3,knight2011internet} and stochastic algorithm-based routing design \cite{yang2016towards,rabjerg2021exploiting,zhao2021multi_4,8068282_5}.

\par
In the literature where the impact of communication distance on network performance is not ignored, the graph theory-based routing studies are based on deterministic network topologies \cite{9348676_1,shen2020dynamic_2} or refer to existing deterministic network models \cite{geng2021agent_3,knight2011internet}. However, since LEO satellites are not geosynchronous, these methods can only provide accurate performance analysis results for deterministic constellations at specific times. For example, the performance evaluation results obtained at a certain moment may no longer be applicable due to changes in distance caused by topology changes at the next moment.

\par
Conversely, stochastic algorithms meet the needs of routing algorithm design in dynamic topologies but are not able to provide analytical expressions for performance metrics \cite{yang2016towards,rabjerg2021exploiting}. The main reason is that stochastic algorithms, such as the ant colony algorithm \cite{zhao2021multi_4} and particle swarm optimization algorithm \cite{8068282_5}, exhibit random performance. If we aim to compare the performance of STR and ISR by stochastic algorithm design, a large number of algorithm implementations should be conducted to obtain the average performance of energy efficiency. Furthermore, every time parameters like the number of satellites or altitude change, the algorithm requires re-execution and the process is computationally expensive. 

\par
Based on the above discussion, we need to find a mathematical tool suitable for low-complexity dynamic network performance analysis. The stochastic geometry (SG) analytical framework simulates dynamic topologies by randomly modeling satellite positions \cite{Al-1}. {\color{black} The core idea of the SG framework is to trade modeling accuracy for analytical tractability. With analytical tractability, metrics are expressed as functions of system parameters such as the number of satellites, significantly reducing computational complexity \cite{ok-2}. In the SG model, LEO satellites are assumed to be randomly distributed at any location, which differs from the deterministic orbital motion with fixed inclinations observed in reality.

\par
Fortunately, numerous studies have demonstrated that the impact of such unrealistic spatial distributions on performance evaluation results is minimal. In \cite{yastrebova2020theoretical}, the authors proved the accuracy of the spherical Poisson point process (PPP) in interference analysis by comparing the estimated performance with those obtained from STK software simulations. In \cite{ok-1}, the authors showed that the coverage probability estimated by the spherical binomial point process (BPP) model is identical to the results obtained from the deterministic Walker constellation configuration. It is worth mentioning that the spherical binomial point process (BPP) is one of the most common models for LEO satellite constellations \cite{talgat2020stochastic}. The authors in \cite{al-3} and \cite{al2021session} also conducted a comparison with the Walker configuration and demonstrated that the pass duration calculated using the spherical SG framework is accurate. In addition, authors in \cite{242206} proved that the difference in performance estimated by the SG-based models and the Starlink constellation is minimal. Given the framework's strong analytical tractability, a significant body of literature has adopted the SG framework to model large-scale LEO satellite constellations and analyze their performance metrics.

\par
So far, the research content addressed in this article has not been discussed in other literature. The research that is closest to this article is our previous studies \cite{wang2022stochastic} and \cite{wang2023reliability}, which have developed satellite routing based on SG. However, neither study considered the channel model, hence the inability to analyze energy efficiency. To comprehensively leverage the advantages of the SG framework in analyzing both topological randomness and channel randomness, models for satellite-to-ground and inter-satellite links need to be integrated into satellite routing. }

\subsection{Contribution}\label{secI-B}
As the pioneering study to integrate channel models into SG-based satellite routing, the contributions of this article are as follows.

\begin{itemize}
    \item To quantitatively compare the communication costs of STR and ISR, we introduce energy efficiency, which is defined as the ratio of the data transmitted to the energy consumed \cite{yang2016towards}, and price ratio factor, which represents the ratio between the price of energy in space and that on the ground. This is the first study of energy efficiency in satellite communications under the SG framework.  
    \item We formulate an optimization problem under an ideal scenario, where satellites are available at any location, to maximize energy efficiency.    
    The solution to this problem provides upper bounds for the energy efficiencies of STR and ISR.
    % In an ideal scenario where satellites are available at any location, optimization problems aiming to maximize energy efficiency are formulated. 
    % By solving these problems, we obtain upper bounds for the energy efficiencies of STR and ISR.
    \item We design routing relay selection strategies for STR and ISR. Note that the strategies are suitable for arbitrary LEO constellation configuration/topology, and thus have a wider applicability than existing methods. 
    \item Numerical results show that in terms of energy efficiency, the proposed strategies offer significant advantages compared to other existing routing methods and can approach ideal upper bounds. 
    \item Furthermore, analytical expressions of availability probability and energy efficiency for proposed algorithms are derived. In addition, the energy efficiency of STR and ISR are compared with different price ratio factors. 
\end{itemize}

\section{System Model}
In this section, we first introduce the spatial distribution models of relay GWs and satellites, and the topology model of routing. Subsequently, we establish channel models for both STLs and ISLs. 
Finally, we present the formal definition of energy efficiency.

\subsection{Spatial Configuration}
We consider a communication system comprising $N_g$ ground GWs and $N_s$ satellites. 
The GWs are located on a sphere with a radius $R_{\oplus}$, which is the radius of the Earth, forming a spherical homogeneous BPP denoted as $\mathcal{X}=\{x_1,x_2,...,x_{N_g}\}$, where $x_i$ is the location of the $i^{th}$ GW.
The satellites are distributed on a sphere around the Earth with a radius $R_s=R_{\oplus}+h_s$, where $h_s$ represents the altitude of the satellites.
These $N_s$ satellites form an independent spherical homogeneous BPP, denoted as $\mathcal{Y}=\{y_1,y_2,...,y_{N_s}\}$, where $y_i$ is the location of the $i^{th}$ satellite.

\begin{figure}[ht]
\centering
\includegraphics[width=0.9\linewidth]{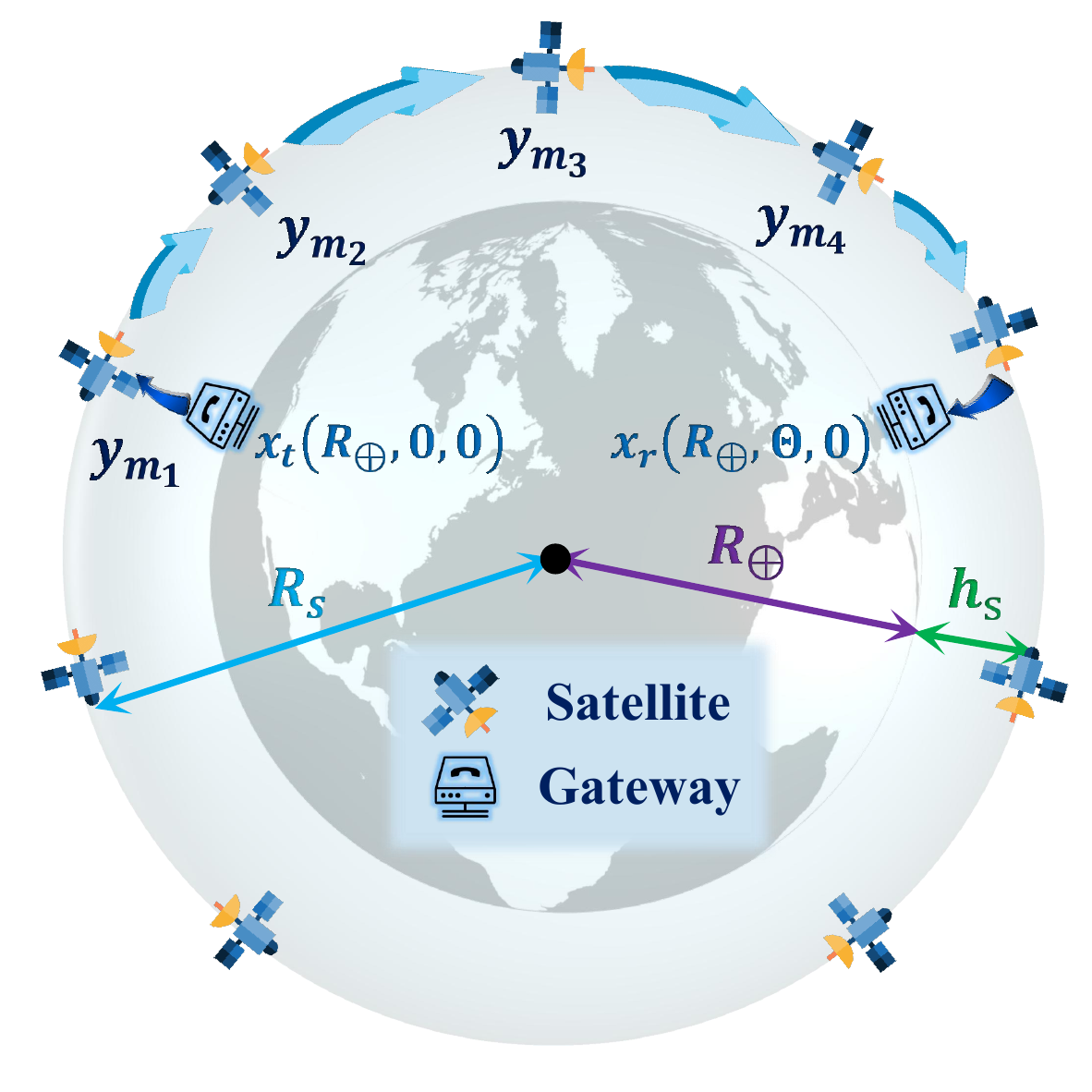}
\caption{A schematic diagram of the spatial configuration.}
\label{figure1-2}
\end{figure}

\par
% As for the modeling of routing, 
We consider a route to be initiated by a ground transmitter and pass through multiple relays before reaching the intended receiver on Earth. 
According to Slivnyak's theorem \cite{feller1991introduction}, the rotation of the coordinate system does not affect the distribution of homogeneous point processes. Without loss of generality, we take the Earth center as the origin. In the spherical coordinate system, the coordinates of the transmitter and receiver are denoted as $x_t (R_{\oplus},0,0)$ and $x_r (R_{\oplus},\Theta,0)$, respectively;  {\color{black}As shown in Fig.~\ref{figure1-2},} $\Theta$ represents the polar angle of the receiver, while the azimuth angle of the receiver is $0$. The relationship between $\Theta$ and the Euclidean distance between the transmitter and receiver $D_{t,r}$ is given by $\Theta = 2 \arcsin \left( 
\frac{D_{t,r}}{2 R_{\oplus}} \right)$.

\par
In this article, we refer to the signal transmission from one communication device to another as a hop.
For a route composed of $N_Q$ hops, where $Q \in \{{\mathrm{ISR}}, {\mathrm{STR}}\}$, we denote it by $\mathcal{M}_{| N_Q} = \{m_1,..., m_{N_Q - 1}\}$, in which $m_{i-1}$ and $m_i$ are the indices of the relays corresponding to the $i^{th}$ hop, $2 \leq i \leq N_Q - 1$. {\color{black}An example of ISR is provided in Fig.~\ref{figure1-2}.}
% A single signal transmission from one communication device to another is referred to as a hop. Given the number of hops is $N_Q$, a route can be expressed as $\mathcal{M}_{| N_Q} = \{m_1,..., m_{N_Q - 1}\}$, where $Q = \{{\mathrm{ISR}}, {\mathrm{STR}}\}$. $m_{i-1}$ and $m_i$ are the indices of the relays corresponding to the $i^{th}$ hop, $2 \leq i \leq N_Q - 1$. 
The first hop of the route occurs from the ground transmitter located at $x_t$ to the satellite located at $y_{m_1}$, while the final hop takes place from the satellite located at $y_{m_{N_Q - 1}}$ to the ground receiver at $x_r$, $Q = \{{\mathrm{ISR}}, {\mathrm{STR}}\}$. 
For ISRs, all relays are satellites, i.e., $m_i,\, \forall \, 2 \leq i \leq N_Q - 1$, are the indices of satellites. 
In contrast, relays in STR consist of satellites and GWs. 
Specifically, when $i$ is odd, $m_i$ represents the index of a satellite; otherwise, $m_i$ represents the index of a GW. In addition, $N_{\mathrm{STR}}$ is required to be an odd number.

\subsection{Channel Model}
% Channel fading models vary between STLs and ISLs due to distinct communication environments. 
The characteristic of channel fading differs between STLs and ISLs due to the distinct communication environments. 
Based on existing research, we consider the channel fading model to follow the free space fading model experienced with large-scale fading and small-scale fading, i.e., 
\begin{equation}
    H_Q (l) = \left( \frac{\lambda_Q}{4\pi l} \right)^2 \zeta_Q W_Q,
\end{equation}
where $l$ is the Euclidean distance between the transmitter and receiver, $\lambda_Q$, $\zeta_Q$, and $W_Q$ denote the wavelength, additional attenuation during propagation in the air, and the power of small-scale fading, respectively; the link index $Q$ is ${\rm{ST}}$ for STLs and ${\rm{SS}}$ for ISLs.
Note that for ISLs, the attenuation in the air is negligible; thus, $\zeta_{\rm{SS}}=1$.

\par
Next, we detail the model of the small-scale fading of the STL $W_{\rm{ST}}$ and that of the ISL $W_{\rm{SS}}$. 
The shadowed-Rician (SR) fading is one of the most widely used models depicting small-scale fading in STLs, considering both the shadowing effect and the multi-path effect of the link. 
The cumulative distribution function (CDF) of the SR fading power $W_{\rm{ST}}$ is given as follows \cite{talgat2020stochastic}:
\begin{equation}
\begin{split}
    F_{W_{\rm{ST}}} (w) & = \left( \frac{2 b_0 n_0 \mathscr{m}}{2 b_0 n_0 + \Omega} \right)^{n_0} \sum_{z=0}^{\infty} \frac{(n_0)_z}{z! \Gamma(z+1)} \\
    & \times \left( \frac{\Omega}{2 b_0 n_0 + \Omega} \right)^z \Gamma_l\left(z+1, \frac{w}{2b_0} \right),
\end{split}
\end{equation}
where $(n_0)_z$ is the Pochhammer symbol, $n_0$, $b_0$, and $\Omega$ are intrinsic parameters of the SR fading, while $\Gamma( \cdot )$ and $\Gamma_l(\cdot, \cdot )$ denote the gamma function and lower incomplete gamma function, respectively.

\par
% Considering that 
Unlike terrestrial environments, the shadowing and multi-path effects in the space environment are generally negligible.
Moreover, the impact of atmospheric turbulence on ISLs can also be neglected \cite{gopal2014modulation}, while the pointing error becomes the primary factor determining the distribution of small-scale fading. 
Therefore, we consider $W_{\rm{SS}}$ follows the pointing error model given in \cite{ata2022performance}. 
Given the deviation angle of the beam, $\theta_d$, the conditional probability density function (PDF) of the pointing error gain $W_{\rm{SS}}$ can be written as\footnote{The vertical and horizontal deviations are presumed to exhibit the same jitter variances, illustrating the scenario of peak pointing error.}
\begin{equation}
\label{pointing_error}
    f_{W_{\rm{SS}} \, | \, \theta_d}\left ( w \right ) = \frac{\eta_s^2 w^{\eta_s^2-1 } \cos\left ( \theta_d \right )}{A_0^{\eta_s^2}},  \ 0 \leq w \leq A_0,
\end{equation}
where $\eta_s$ and $A_0$ are parameters of the pointing error, and the deviation angle $\theta_d$ is a random variable that follows the Rayleigh distribution with variance $\varsigma^2$, i.e.,
\begin{equation}
    f_{\theta_d}\left ( \theta_d \right )=\frac{\theta_d}{\varsigma^2}\exp\left ( -\frac{\theta_d^2}{2\varsigma^2} \right ), \ \theta_d\geq 0.
\end{equation}

\par
% Based on the above channel fading models, the received power can be categorized into the following three cases:
Across the multiple hops in a route, the signal power received at a node can be categorized into the following three cases:
\begin{itemize}
    \item $(c_1)$: The signal is transmitted by either a GW or the ground transmitter and received by a satellite. It corresponds to the odd hops in STR and the first hop in ISR. 
    \item $(c_2)$: The signal is transmitted by a satellite and received by either a GW or the ground receiver. It corresponds to the even hops in STR and the last hop in ISR. 
    \item $(c_3)$: The signal is transmitted by a satellite and received by a satellite. It corresponds to the intermediate hops in ISR (all the hops except the first and last one). 
\end{itemize}
Mathematically, the received power can be respectively expressed for the cases $(c_1)$, $(c_2)$, and $(c_3)$ as follows: 
\begin{align}\label{rho_i}
    \left\{
 	\begin{array}{lll}
    \rho_r^{(1)} (l) = \rho_t^{(1)} G_{\rm{ST}} H_{\mathrm{ST}} (l), & l \leq l_{\max}^{(1)}, \\
    \rho_r^{(2)} (l) = \rho_t^{(2)} G_{\rm{ST}} H_{\mathrm{ST}} (l), & l \leq l_{\max}^{(2)}, \\
    \rho_r^{(3)} (l) = \rho_t^{(3)} G_{\rm{SS}} H_{\mathrm{SS}} (l), & l \leq l_{\max}^{(3)}, 
	\end{array}
	\right.
	%\vspace{-0.15cm}
\end{align}
where $\rho_t^{(1)}$, $\rho_t^{(2)}$, and $\rho_t^{(3)}$ denote the corresponding transmission power. $G_{\rm{ST}}$ represents the bidirectional antenna gain between the satellite and the GW in the STL, while $G_{\rm{SS}}$ refers to the bidirectional antenna gain between the transmitting and receiving satellites in the ISL. $l_{\max}^{(1)}$, $l_{\max}^{(2)}$ and $l_{\max}^{(3)}$ are the maximum distances that can maintain stable communication for communication devices. To ensure links are not blocked by the Earth, the conditions $l_{\max}^{(1)} \leq \sqrt{R_s^2 - R_{\oplus}^2}$, $l_{\max}^{(2)} \leq \sqrt{R_s^2 - R_{\oplus}^2}$, and $l_{\max}^{(3)} \leq 2\sqrt{R_s^2 - R_{\oplus}^2}$ need to be satisfied.

\subsection{Performance Metric}
This subsection provides the definition of the routing energy efficiency and single-hop energy efficiency for STR and ISR. Taking into account the energy price difference between space and ground, we weighted the energy consumed on satellites by the price ratio factor $\beta>1$, which represents the ratio between the price of energy in space and that on the ground. 

\begin{definition}[Routing Energy Efficiency] \label{defenergy}
    Energy efficiency is defined as the ratio of the amount of data transmitted to the weighted energy consumed in the routing. The latter is the sum of the GWs' transmission power, added to the sum of satellites' transmission power weighted by the price ratio factor $\beta$.
\end{definition}
We provide the mathematical expression for the routing energy efficiency of STR, while the expression for ISR will be presented later. Denoting the data packet size as $\xi$ and the time required to transmit the data packet for the $i^{th}$ hop as $T_{{\mathrm{STR}}, i}$, the routing energy efficiency is
\begin{equation}
\begin{split}
    E_{\mathrm{rout}}^{\mathrm{STR}} & = \frac{\xi}{\sum_{i=1}^{N_{\mathrm{STR}}} T_{{\mathrm{STR}},i} \, \rho_{t,i}^{\mathrm{STR}}} = \frac{\xi}{\sum_{i=1}^{N_{\mathrm{STR}}} \frac{\xi}{\mathcal{R}_{{\mathrm{STR}},i}} \, \rho_{t,i}^{\mathrm{STR}}} \\
    & = \left( \sum_{i=1}^{N_{\mathrm{STR}}} \frac{\rho_{t,i}^{\mathrm{STR}}}{\mathcal{R}_{{\mathrm{STR}},i}} \right)^{-1} = \left( \sum_{i=1}^{N_{\mathrm{STR}}} \frac{1}{E_{{\mathrm{STR}},i}} \right)^{-1},
\end{split}
\end{equation}
where $\rho_{t,i}^{\mathrm{STR}}$, $\mathcal{R}_{{\mathrm{STR}},i}$, and $E_{{\mathrm{STR}},i} = \mathcal{R}_{{\mathrm{STR}},i} / \rho_{t,i}^{\mathrm{STR}}$ denote the transmission power, achievable data rate, and energy efficiency of the $i^{th}$ hop, respectively. 
The achievable data rate $\mathcal{R}_{{\mathrm{STR}},i}$ is defined as the ergodic capacity from the Shannon-Hartley theorem over a fading communication link \cite{wang2022ultra}. 
Using the expression of $\mathcal{R}_{{\mathrm{STR}},i}$, the energy efficiency for the $i^{th}$ hop can be written as
\begin{align}\label{ESTR}
\begin{small}
    E_{{\mathrm{STR}},i} =
    \left\{
 	\begin{array}{lll}
    \frac{1}{\rho_t^{(1)}} B_{\rm{ST}} \log_2 \left( 1 + \frac{\rho_r^{(1)} (l_i)}{\sigma_s^2} \right), & {i \mathrm{\ is \, odd}}, \\
    \frac{1}{\beta \, \rho_t^{(2)}} B_{\rm{ST}} \log_2 \left( 1 + \frac{\rho_r^{(2)} (l_i)}{\sigma_g^2} \right), & {i \mathrm{\ is \, even}},
	\end{array}
	\right.
\end{small}
\end{align}
where $l_i$ is the Euclidean distance of the $i^{th}$ hop and $B_{\rm{ST}}$ denotes the bandwidth of STL. $\sigma_g^2$ represents the noise power at the ground receiver or the GW, while $\sigma_s^2$ denotes the noise power at the satellite. 

\par
Similarly, the routing energy efficiency of ISR is given as $E_{\mathrm{rout}}^{\mathrm{ISR}} = \left( \sum_{i=1}^{N_{\mathrm{ISR}}} \frac{1}{E_{{\mathrm{ISR}},i}} \right)^{-1}$, where $E_{{\mathrm{ISR}},i}$ is the energy efficiency for the $i^{th}$ hop in ISR. We denote the bandwidth of ISL as $B_{\rm{SS}}$, and $E_{{\mathrm{ISR}},i}$ can be formally written as
\begin{align}\label{EISR}
\begin{small}
     \! \! \! E_{{\mathrm{ISR}},i} = \left\{
 	\begin{array}{lll}
    \frac{1}{\rho_t^{(1)}} B_{\rm{ST}} \log_2 \left( 1 + \frac{\rho_r^{(1)} (l_i)}{\sigma_s^2} \right), & i = 1, \\
    \frac{1}{\beta \, \rho_t^{(2)}} B_{\rm{ST}} \log_2 \left( 1 + \frac{\rho_r^{(2)} (l_i)}{\sigma_g^2} \right), & i = N_{\mathrm{ISR}}, \\
    \frac{1}{\beta \, \rho_t^{(3)}} B_{\rm{SS}} \log_2 \left( 1 + \frac{\rho_r^{(3)} (l_i)}{\sigma_s^2} \right), & \! \! \! \! \! 1<i<N_{\mathrm{ISR}}.
	\end{array}
	\right.
\end{small}
\end{align}

{\color{black} It is worth noting that, considering routing differs from communication coverage scenarios, we do not take into account the impact of interference on energy efficiency. Firstly, satellites typically select a specific satellite or GW as the next hop, which constitutes point-to-point communication. In contrast, coverage involves one-to-many communications, where a satellite simultaneously provides coverage for multiple devices within the beam's range, making interference significant. Secondly, the deployment density of satellites or GWs is generally lower than the user density receiving coverage from the satellite. For example, consider a constellation with $1000$ satellites at an altitude of $500$ km. Even assuming all satellites are oriented toward the Earth's center (maximizing beam coverage) and equipped with wide beams with a central angle of $\pi/6$, the probability of having more than two GWs within the coverage area is only $0.57\%$. Given that the actual beam orientation is more random and beams are narrower, the probability of a GW receiving interference from other satellites is minimal, making it easy to filter out interference signals.}

\section{Routing Algorithm Design and Performance Analysis}
In this section, we introduce an optimization problem to maximize routing energy efficiency by adequately selecting satellites as relays. 
As this problem is non-convex, we propose two algorithms providing relay selection schemes for STR and ISR, respectively. 
We analyze the performance of these schemes in terms of system-level metrics such as availability and energy efficiency. 
We also present a few technical lemmas to facilitate the algorithm design as well as the derivation of performance metrics. 
% Additionally, several lemmas are presented to facilitate algorithm design and the derivation of performance metrics. 

\subsection{Problem Formulation}
This subsection aims to maximize the routing energy efficiency of STR $E_{\mathrm{rout}}^{\mathrm{STR}}$ by optimizing the number of hops $N_{\mathrm{STR}}$ and relay positions selection $\mathcal{M}_{|N_{\mathrm{STR}}}$. 
Note that maximizing $E_{\mathrm{rout}}^{\mathrm{STR}}$ is equivalent to minimizing its reciprocal, which leads to the following optimization problem:
\begin{equation} 
	\mathscr{P}_{{\mathrm{STR}},1}: \ \underset{N_{\mathrm{STR}}, \, \mathcal{M}_{|N_{\mathrm{STR}}} }{\mathrm{minimize}} \  \frac{1}{E_{\mathrm{rout}}^{\mathrm{STR}}} = \sum_{i=1}^{N_{\mathrm{STR}}} \frac{1}{E_{{\mathrm{STR}},i}}.
\end{equation}
The optimization objective of the above problem cannot be explicitly expressed in terms of the decision variables, namely $N_{\mathrm{STR}}$ and $\mathcal{M}_{|N_{\mathrm{STR}}}$. Therefore, $\mathscr{P}_{\mathrm{STR}}$ needs to be transformed into a more manageable form. 
Before proceeding, we need to first introduce the below definition.

\begin{definition}[Central Angle]
    Given two communication devices $x_i$ and $y_j$, the central angle between them is the angle formed by the line connecting the center of the Earth to $x_i$ and the line connecting the center of the Earth to $y_j$.
\end{definition}

\par
Based on this definition, we propose a proposition to improve the tractability of $\mathscr{P}_{\mathrm{STR}}$.

\begin{proposition}\label{propsition1}
    Given the number of hops $N_{\mathrm{STR}}$, the ideal relay positions in the spherical coordinate system of STR are
\begin{equation}
    \begin{split}
    & \bigg\{ (R_s, \theta_{{\mathrm{STR}},1}, 0), (R_{\oplus}, \theta_{{\mathrm{STR}},1}+\theta_{{\mathrm{STR}},2}, 0), \\
    & (R_s, 2\theta_{{\mathrm{STR}},1}+\theta_{{\mathrm{STR}},2}, 0), (R_{\oplus}, 2\theta_{{\mathrm{STR}},1} + 2 \theta_{{\mathrm{STR}},2}, 0), \\ 
    &  \dots, 
    %\left( R_{\oplus}, \frac{N_{\mathrm{STR}}}{2} \theta_{{\mathrm{STR}},1} + \left( \frac{N_{\mathrm{STR}}}{2}-1 \right) \theta_{{\mathrm{STR}},2}, 0 \right), \\
    \left( R_s, \frac{N_{\mathrm{STR}}}{2} \theta_{{\mathrm{STR}},1} + \frac{N_{\mathrm{STR}}}{2} \theta_{{\mathrm{STR}},2}, 0 \right) \bigg\},
    \end{split}
\end{equation}
where $\theta_{{\mathrm{STR}},1}$ and $\theta_{{\mathrm{STR}},2}$ represent the central angles of odd hops and even hops in STR, respectively, satisfying $\Theta = \frac{N_{\mathrm{STR}}}{2} (\theta_{{\mathrm{STR}},1} + \theta_{{\mathrm{STR}},2})$.
\end{proposition}
\begin{IEEEproof}
See Appendix~\ref{app:propsition1}.
\end{IEEEproof}

The above proposition indicates that the ideal relay positions can be uniquely determined only when the values of $N_{\mathrm{STR}}$ and $\theta_{{\mathrm{STR}},1}$ are specified. Given the ideal relay positions, we can search for the relays closest to these positions to act as the relays. 

In a similar vein, we present the optimization problem for ISL and its corresponding proposition as follows:
\begin{equation} 
	\mathscr{P}_{{\mathrm{ISR}},1}: \ \underset{N_{\mathrm{ISR}}, \, \mathcal{M}_{|N_{\mathrm{ISR}}} }{\mathrm{minimize}} \  \frac{1}{E_{\mathrm{rout}}^{\mathrm{ISR}}} = \sum_{i=1}^{N_{\mathrm{ISR}}} \frac{1}{E_{{\mathrm{ISR}},i}}.
\end{equation}
\begin{proposition}\label{propsition2}
    Given the number of hops $N_{\mathrm{ISR}}$, the ideal relay positions in the spherical coordinate system of ISL are
\begin{sequation}
    \begin{split}
    & \{ (R_s, \theta_{{\mathrm{ISR}},1}, 0), (R_s, \theta_{{\mathrm{ISR}},1}+\theta_{{\mathrm{ISR}},3}, 0), (R_s, \theta_{{\mathrm{ISR}},1} + 2\theta_{{\mathrm{ISR}},3},0), \\ 
    & \dots, \left( R_s, \theta_{{\mathrm{ISR}},1} + \left( N_{\mathrm{ISR}}-2 \right) \theta_{{\mathrm{ISR}},3}, 0 \right) \},
    \end{split}
\end{sequation}
where $\theta_{{\mathrm{ISR}},1}$, $\theta_{{\mathrm{ISR}},2}$, and $\theta_{{\mathrm{ISR}},3}$ stand for the central angles of the first hop, last hop, and middle hops in ISR, respectively, satisfying $\theta_{{\mathrm{ISR}},1} + \left( N_{\mathrm{ISR}}-2 \right) \theta_{{\mathrm{ISR}},3} + \theta_{{\mathrm{ISR}},2} = \Theta$.
\end{proposition}
\begin{IEEEproof}
The proof of Proposition~\ref{propsition2} is similar to that of Proposition~\ref{propsition1}, therefore omitted here. 
\end{IEEEproof}

In this case, the ideal relay positions in ISR are dependent on the specific values of $N_{\mathrm{ISR}}$, $\theta_{{\mathrm{ISR}},1}$, and $\theta_{{\mathrm{ISR}},2}$. 
It can be inferred that the algorithm's computational complexity for ISR is higher than that for STR, given that the former has one more degree of freedom.

\subsection{Preliminaries}
In this subsection, we provide preliminaries for algorithm design, laying the groundwork for the subsequent algorithm development. 

\par
According to Proposition \ref{propsition1} and \ref{propsition2}, the ideal relay positions of STR and ISR depend on the decision variables $\{N_{\mathrm{STR}}, \theta_{{\mathrm{STR}},1}\}$ and $\{N_{\mathrm{ISR}}, \theta_{{\mathrm{ISR}},1}, \theta_{{\mathrm{ISR}},2} \}$, respectively. 
Correspondingly, we derive the analytical expressions of the energy efficiency under the two sets of decision variables. 

% depends on decision variables $\{N_{\mathrm{STR}}, \theta_{{\mathrm{STR}},1}\}$, while that of ISR depends on decision variables $\{N_{\mathrm{ISR}}, \theta_{{\mathrm{ISR}},1}, \theta_{{\mathrm{ISR}},2} \}$. 
% Therefore, we derive the energy efficiency to be optimized for the two sets of decision variables. 
As the small-scale fading in the objective function is a random variable, we take the expectation to facilitate optimization.

\par
\begin{lemma}\label{lemma1}
Given that the central angle of a certain hop is $\theta$, the average energy efficiency of this hop is given in (\ref{Ehop}) at the top of the next page.
\begin{table*}
\begin{align}\label{Ehop}
    \left\{
 	\begin{array}{lll}
    \overline{E}_{\mathrm{hop}}^{(1)} (\theta) = \frac{1}{\rho_t^{(1)}} B_{\rm{ST}} \log_2 \left( 1 +  \frac{\rho_t^{(1)} \zeta_{\rm{ST}} G_{\rm{ST}} (2b_0 + \Omega)}{ \left( R_{\oplus}^2 + R_s^2 - 2R_{\oplus}R_s \cos\theta \right) \sigma_s^2 } \left( \frac{\lambda_{\rm{ST}}}{4\pi} \right)^2 \right), & {\mathrm{for \ } (c_1)}, \\
    \overline{E}_{\mathrm{hop}}^{(2)} (\theta) = \frac{1}{\beta \, \rho_t^{(2)}} B_{\rm{ST}} \log_2 \left( 1 +  \frac{\rho_t^{(2)} \zeta_{\rm{ST}} G_{\rm{ST}} (2b_0 + \Omega)}{ \left( R_{\oplus}^2 + R_s^2 - 2R_{\oplus}R_s \cos\theta \right) \sigma_g^2 } \left( \frac{\lambda_{\rm{ST}}}{4\pi} \right)^2 \right), & {\mathrm{for \ } (c_2)}, \\
    \overline{E}_{\mathrm{hop}}^{(3)} (\theta) = \frac{1}{\beta \, \rho_t^{(3)}} B_{\rm{SS}} \log_2 \left( 1 + \rho_t^{(3)} \zeta_{\rm{SS}} G_{\rm{SS}} \left( \frac{\lambda_{\rm{SS}}}{8\pi R_s \sin(\theta/2)} \right)^2 \frac{A_0 \eta_s^2}{1 + \eta_s^2} \left( 1 - \varsigma^2 \right) \sigma_s^{-2} \right),  & {\mathrm{for \ } (c_3)}.
	\end{array}
	\right.
\end{align}
\hrule
\end{table*}
\begin{IEEEproof}
    See Appendix~\ref{app:lemma1}.
\end{IEEEproof}
\end{lemma}

\par
Using Lemma~\ref{lemma1}, we can rewrite the optimization problems $\mathscr{P}_{{\mathrm{STR}},1}$ and $\mathscr{P}_{{\mathrm{ISR}},1}$ into a solvable format. The objective functions in both revised optimization problems $\mathscr{P}_{{\mathrm{STR}},2}$ and $\mathscr{P}_{{\mathrm{ISR}},2}$ can be explicitly defined by the decision variables. For STR, we denote the optimization problem as: \par $\mathscr{P}_{{\mathrm{STR}},2} :$
\begin{subequations} 
\begin{alignat}{2}
%\mathscr{P}_{{\mathrm{STR}},2}: \ E_{\mathrm{opt}}^{\mathrm{STR}} =  
\underset{N_{\mathrm{STR}}, \, \theta_{{\mathrm{STR}},1} }{\mathrm{maximize}} \ \ \ &  \left( \frac{N_{\mathrm{STR}}}{2 \, \overline{E}_{\mathrm{hop}}^{(1)} (\theta_{{\mathrm{STR}},1})} + \frac{N_{\mathrm{STR}}}{2 \, \overline{E}_{\mathrm{hop}}^{(2)} (\theta_{{\mathrm{STR}},2}) } \right)^{-1} \label{opt3-0} \\
\mathrm{subject \ to} \ \ \  & \ \frac{N_{\mathrm{STR}}}{2} (\theta_{{\mathrm{STR}},1} + \theta_{{\mathrm{STR}},2}) = \Theta, \label{opt3-1} \\
& \ \theta_{{\mathrm{STR}},Q} \leq \arccos\left( \frac{R_s^2 + R_{\oplus}^2 - \big( l_{\max}^{(Q)} \big)^2}{ 2 R_s R_{\oplus}} \right), \label{opt3-2}
\end{alignat}
\end{subequations}
where $Q=\{1,2\}$, constraint (\ref{opt3-1}) is based on Proposition~\ref{propsition1} and constraint (\ref{opt3-2}) ensures that the distance for each hop does not exceed the maximum distance. For ISR, we denote the optimization problem as: \par $\mathscr{P}_{{\mathrm{ISR}},2}$:
\begin{footnotesize}
\begin{subequations} 
\begin{alignat}{2}  
\underset{N_{\mathrm{ISR}}, \, \theta_{{\mathrm{ISR}},1}, \, \theta_{{\mathrm{ISR}},2} }{\mathrm{maximize}} \ & \! \! \! \! \! \left( \frac{1}{ \overline{E}_{\mathrm{hop}}^{(1)} (\theta_{{\mathrm{ISR}},1})} + \frac{N_{\mathrm{ISR}}-2}{ \overline{E}_{\mathrm{hop}}^{(3)} (\theta_{{\mathrm{ISR}},3}) } + \frac{1}{ \overline{E}_{\mathrm{hop}}^{(2)} (\theta_{{\mathrm{ISR}},2}) } \right)^{-1} \label{opt4-0} \\
\mathrm{subject \ to} \ \ \  & \  \theta_{{\mathrm{STR}},1} + \theta_{{\mathrm{STR}},2} + (N_{\mathrm{ISR}}-2) \, \theta_{{\mathrm{STR}},3} = \Theta, \label{opt4-1} \\
& \! \! \! \! \!  \! \! \! \! \! \theta_{{\mathrm{ISR}},Q} \leq \arccos\left( \frac{R_s^2 + R_{\oplus}^2 - \big( l_{\max}^{(Q)} \big)^2}{ 2 R_s R_{\oplus}} \right), \, Q=\{1,2\}, \label{opt4-2} \\
& \ \theta_{{\mathrm{ISR}},3} \leq 2 \arcsin\left( \frac{ l_{\max}^{(3)} }{2R_s} \right), \label{opt4-3}
\end{alignat}
\end{subequations}
\end{footnotesize}
where constraint (\ref{opt4-1}) is based on Proposition~\ref{propsition2}. Constraints (\ref{opt4-2}) and (\ref{opt4-3}) stipulate that the distance of the first/last and middle hops do not exceed the maximum distances.

\begin{proposition}\label{proposition3}
The methods of directly solving $\mathscr{P}_{{\mathrm{STR}},2}$ and $\mathscr{P}_{{\mathrm{ISR}},2}$ to design STR and ISR are referred to as the ideal scenario solutions. The corresponding optimal energy efficiencies $E_{\mathrm{opt}}^{\mathrm{STR}}$ and  $E_{\mathrm{opt}}^{\mathrm{ISR}}$ serve as ideal upper bounds on energy efficiency that cannot be attained, where $E_{\mathrm{opt}}^{\mathrm{STR}}$ and  $E_{\mathrm{opt}}^{\mathrm{ISR}}$ are objective functions defined in (\ref{opt3-0}) and (\ref{opt4-0}).
\end{proposition}
\begin{IEEEproof}
Note that constraints (\ref{opt3-1}) and (\ref{opt4-1}) can only be satisfied when the relays are placed exactly at the ideal positions specified in Proposition~\ref{propsition1} and Proposition~\ref{propsition2}. However, this is unattainable in practical constellations, thus $E_{\mathrm{opt}}^{\mathrm{STR}}$ and  $E_{\mathrm{opt}}^{\mathrm{ISR}}$ are ideal upper bounds on energy efficiency.
\end{IEEEproof}

Due to the manageable computational complexity of the ideal scenario solution, in subsequent numerical results, we achieve it through an exhaustive search of the decision variables. So far, routing designs are based on the assumption that relays are available anywhere. This assumption is strong and unrealistic. Therefore, we present the following lemmas and algorithms to address this issue.
However, when relay device location models follow BPPs, their positions may deviate from the ideal positions given in Proposition~\ref{propsition1} and Proposition~\ref{propsition2}, leading to an increase in the average distance for each hop. We use the distance scaling factor to characterize the increase in length. 

\begin{definition}[Distance Scaling Factor]
In the scenario where two ideal relay positions are given, two communication devices closest to each respective relay position can be identified from the BPP (or BPPs). The distance scaling factor is the ratio of the average Euclidean distance between these two devices to the Euclidean distance of two ideal relay positions. 
\end{definition}

The distance scaling factor is influenced by various factors, e.g., the distance between relay positions and the number of devices in the BPP. The following lemma provides a specific expression.

\begin{lemma}\label{lemma2}
Given that the central angle of a hop is $\theta$. When the hop is the first or last hop in STR or ISR, the distance scaling factor  is
\begin{sequation}\label{alpha1}
\begin{split}
    \alpha^{(1)} \left( \theta \right) & = \frac{N_s}{4 \pi \sqrt{R_{\oplus}^2 + R_s^2 - 2 R_{\oplus} R_s \cos\theta}} \int_0^{2\pi} \int_0^{\pi} \sin\psi \\
    & \times \left( \frac{ 1 + \cos\psi }{2} \right)^{N_s-1} d \left( R_s,\psi,\varphi; R_{\oplus},\theta,0 \right) \mathrm{d}\psi \mathrm{d}\varphi,
\end{split}
\end{sequation}where $d \left( R_1, \theta_1, \varphi_1 ; R_2, \theta_2, \varphi_2 \right )$ in (\ref{alpha1}) represents the Euclidean distance between $(R_1, \theta_1, \varphi_1)$ and $(R_2, \theta_2, \varphi_2)$:
\begin{sequation}\label{d_12}
\begin{split}
    & d \left( R_1, \theta_1, \varphi_1 ; R_2, \theta_2, \varphi_2 \right ) \\
    & =\! \sqrt{R_1^2 \!+\! R_2^2 - 2  R_1 R_2 (\sin\theta_1 \sin\theta_2 \cos(\varphi_1-\varphi_2) \!+\! \cos\theta_1 \cos\theta_2 )}.
\end{split}
\end{sequation}When the hop is one of the intermediate hops in STR, the distance scaling factor can be approximated as 
\begin{sequation}
\begin{split}
    & \alpha^{(2)} \left(\theta\right) = \frac{N_g \, \alpha^{(1)} \left( \theta \right)}{4 \pi \sqrt{R_{\oplus}^2 \!+\! R_s^2 - 2 R_{\oplus} R_s \cos\theta}}  \int_0^{2\pi} \int_0^{\pi} \sin\psi \\
    & \times \left( \frac{ 1 + \cos\psi }{2} \right)^{N_g-1} d \left( R_{\oplus},\psi,\varphi; R_s,\theta,0 \right) \mathrm{d}\psi \mathrm{d}\varphi.
\end{split}
\end{sequation}When the hop is one of the intermediate hops in ISR, the distance scaling factor can be approximated as
\begin{sequation}
\begin{split}
    & \alpha^{(3)} \left(\theta\right) = \bigg( \frac{N_s}{8 \pi R_s \sin(\theta/2)} \int_0^{2\pi} \int_0^{\pi} \sin\psi \\
    & \times \left( \frac{ 1 + \cos\psi }{2} \right)^{N_s-1} d \left( R_s,\psi,\varphi; R_s,\theta,0 \right) \mathrm{d}\psi \mathrm{d}\varphi \bigg)^2.
\end{split}
\end{sequation}
\end{lemma}
\begin{IEEEproof}
    See Appendix~\ref{app:lemma2}.
\end{IEEEproof}

Based on the above lemma, we propose algorithms to obtain the optimal values of the decision variables in the next subsection when the relay device locations follow BPPs.

\subsection{Routing Algorithms}
This subsection includes the design of two algorithms separately for determining decision variables in STR and ISR. Furthermore, a general relay subset selection algorithm applicable to both STR and ISR is proposed, with its computational complexity analyzed later. 

\begin{algorithm}[!ht] 
    \caption{Exhaustive Search for Optimal Values of $\left\{ N_{\mathrm{STR}}^*, \, \theta_{{\mathrm{STR}},1}^* \right\}$ in STR}
	\label{alg1} 
	\begin{algorithmic} [1]
	\STATE \textbf{Initiate} $N_{\mathrm{STR}}^* \leftarrow 0$, $E_{\mathrm{opt}}^{ \mathrm{STR}} \leftarrow 0$, and $N_{\mathrm{STR}} \leftarrow 2$.

    \STATE $\theta_{\max}^{(Q)} \leftarrow \arccos\left( \frac{R_s^2 + R_{\oplus}^2 - \big( l_{\max}^{(Q)} \big)^2 }{ 2 R_s R_{\oplus} } \right), Q=\{1,2\}$. 
	\WHILE{$N_{\mathrm{STR}} \leq N_{\max}^{\mathrm{STR}}$}
    \STATE $\theta_{{\mathrm{STR}},1} \leftarrow 0$.

    \WHILE{$\theta_{{\mathrm{STR}},1} \leq \min \left\{ \theta_{\max}^{(1)}, \frac{2\Theta}{N_{\mathrm{STR}}} \right\}$}

    \STATE $\theta_{{\mathrm{STR}},2} \leftarrow \frac{2\Theta}{N_{\mathrm{STR}}} - \theta_{{\mathrm{STR}},1}$.
    \STATE $l_{{\mathrm{STR}},Q_1} \leftarrow \sqrt{R_s^2 + R_{\oplus}^2 - 2R_s R_{\oplus} \cos \theta_{{\mathrm{STR}},Q_1}}$, \\ where $Q_1 = \{ 1, 2 \}$.
    \STATE $\widetilde{\theta}_{ {\mathrm{STR} },Q_1}^{(Q_2)} \leftarrow \arccos \left( \frac{R_s^2 + R_{\oplus}^2 - \left( \alpha^{(Q_2)}(\theta_{{\mathrm{STR}},Q_1}) l_{{\mathrm{STR}},Q_1} \right)^2}{2R_s R_{\oplus}} \right)$, \\ where $Q_1 = \{ 1, 2 \}$, $Q_2 = \{ 1, 2 \}$. 
    
	\IF{$\widetilde{\theta}_{ {\mathrm{STR}},1}^{(2)} \leq (1-\varepsilon) \, \theta_{\max}^{(1)}$ and $\widetilde{\theta}_{ {\mathrm{STR}},2}^{(2)} \leq (1-\varepsilon) \, \theta_{\max}^{(2)}$ }
  
    \STATE $E_{\mathrm{rout}}^{\mathrm{STR}} \leftarrow \bigg( \frac{1}{ \overline{E}_{\mathrm{hop}}^{(1)} (\widetilde{\theta}_{{\mathrm{STR}},1}^{(1)})} + \frac{1}{ \overline{E}_{\mathrm{hop}}^{(2)} (\widetilde{\theta}_{{\mathrm{STR}},2}^{(1)}) }$ + \\ \ \ \ \ \ \ \ \ \  $\frac{N_{\mathrm{STR}} - 2}{2 \, \overline{E}_{\mathrm{hop}}^{(1)} (\widetilde{\theta}_{{\mathrm{STR}},1}^{(2)}) } + \frac{N_{\mathrm{STR}} - 2}{2 \, \overline{E}_{\mathrm{hop}}^{(2)} (\widetilde{\theta}_{{\mathrm{STR}},2}^{(2)}) } \bigg)^{-1}$.
        
    \IF{$E_{\mathrm{rout}}^{ \mathrm{STR}} > E_{\mathrm{opt}}^{\mathrm{STR}}$}
    
    \STATE $E_{\mathrm{opt}}^{\mathrm{STR}} \leftarrow E_{\mathrm{rout}}^{ \mathrm{STR}}$, $N_{\mathrm{STR}}^* \leftarrow N_{\mathrm{STR}}$, $\theta_{{\mathrm{STR}},1}^* \leftarrow \theta_{{\mathrm{STR}},1}$.
    \ENDIF
    \ENDIF

    \STATE $\theta_{{\mathrm{STR}},1} \leftarrow \theta_{{\mathrm{STR}},1} + \frac{1}{N_{\mathrm{in}}} \min \left\{ \theta_{\max}^{(1)}, \frac{2\Theta}{N_{\mathrm{STR}}} \right\}$.
    \ENDWHILE
    \STATE $N_{\mathrm{STR}} \leftarrow N_{\mathrm{STR}} + 2$.
	\ENDWHILE
 
	\STATE \textbf{Output}: Optimal values of decision variables $\left\{ N_{\mathrm{STR}}^*, \, \theta_{{\mathrm{STR}},1}^* \right\}$ in STR.
	\end{algorithmic}
\end{algorithm}

\begin{algorithm}[!ht] 
    \caption{Exhaustive Search for Optimal Values of $\left\{ N_{\mathrm{ISR}}^*, \, \theta_{{\mathrm{ISR}},1}^*, \, \theta_{{\mathrm{ISR}},2}^* \right\}$ in ISR}
	\label{alg2} 

	\begin{algorithmic} [1]
	\STATE \textbf{Initiate} $N_{\mathrm{ISR}}^* \leftarrow 0$, $E_{\mathrm{opt}}^{\mathrm{ISR}} \leftarrow 0$, and $N_{\mathrm{ISR}} \leftarrow 2$. 

    \STATE $\theta_{\max}^{(Q)} \leftarrow \arccos\left( \frac{R_s^2 + R_{\oplus}^2 - \left( l_{\max}^{(Q)} \right)^2 }{ 2 R_s R_{\oplus} } \right),Q=\{1,2\}$; $\theta_{\max}^{(3)} \leftarrow 2 \arcsin\left( \frac{ l_{\max}^{(3)} }{ 2 R_s } \right)$.
 
	\WHILE{$N_{\mathrm{ISR}} \leq N_{\max}^{\mathrm{ISR}}$}
    \STATE $\theta_{{\mathrm{ISR}},1} \leftarrow 0$. 

    \WHILE{$\theta_{{\mathrm{ISR}},1} \leq \min \left\{ \theta_{\max}^{(1)}, \Theta \right\}$}

    \STATE $\theta_{{\mathrm{ISR}},2} \leftarrow 0$.
    \WHILE{$\theta_{{\mathrm{ISR}},2} \leq \min \left\{ \theta_{\max}^{(2)}, \Theta - \theta_{{\mathrm{ISR}},1} \right\}$}

    \STATE $\theta_{{\mathrm{ISR}},3} \leftarrow \frac{\Theta - \theta_{{\mathrm{ISR}},1} - \theta_{{\mathrm{ISR}},2}}{N_{\mathrm{STR}}-2}$.
    \STATE $l_{{\mathrm{ISR}},Q_1} \leftarrow \sqrt{R_s^2 + R_{\oplus}^2 - 2R_s R_{\oplus} \cos \theta_{{\mathrm{ISR}} ,Q_1}}$, \\ where $Q_1 = \{ 1, 2 \}$;  $l_{{\mathrm{ISR}},3} \leftarrow 2R_s \sin\left( \frac{\theta_{{\mathrm{ISR}}, 3}}{2} \right)$. 
    \STATE $\widetilde{\theta}_{ {\mathrm{ISR} },Q_1}^{(1)} \leftarrow \arccos \left( \frac{R_s^2 + R_{\oplus}^2 - \left( \alpha^{(1)}(\theta_{{\mathrm{ISR}},Q_1}) l_{{\mathrm{ISR}},Q_1} \right)^2}{2R_s R_{\oplus}} \right)$, \\ where $Q_1 = \{ 1, 2 \}$; \\ $\widetilde{\theta}_{ {\mathrm{ISR} },3}^{(3)} \leftarrow 2 \arcsin \left( \frac{ \alpha^{(3)}(\theta_{{\mathrm{ISR}},3}) l_{{\mathrm{ISR}},3} }{2R_s} \right)$.
    
	\IF{$\widetilde{\theta}_{ {\mathrm{ISR}},1}^{(1)} \leq (1-\varepsilon) \, \theta_{\max}^{(1)}$, $\widetilde{\theta}_{ {\mathrm{ISR}},2}^{(1)} \leq (1-\varepsilon) \, \theta_{\max}^{(2)}$, and $\widetilde{\theta}_{ {\mathrm{ISR}},3}^{(3)} \leq (1-\varepsilon) \, \theta_{\max}^{(3)}$ }
  
    \STATE $E_{\mathrm{rout}}^{\mathrm{ISR}} \leftarrow \bigg( \frac{1}{ \overline{E}_{\mathrm{hop}}^{(1)} (\widetilde{\theta}_{{\mathrm{ISR}},1}^{(1)})} + \frac{1}{ \overline{E}_{\mathrm{hop}}^{(2)} (\widetilde{\theta}_{{\mathrm{ISR}},2}^{(1)}) }$ \\ \ \ \ \ \ \ \ \ \ $ + \frac{N_{\mathrm{ISR}} - 2}{ \overline{E}_{\mathrm{hop}}^{(3)} (\widetilde{\theta}_{{\mathrm{ISR}},3}^{(3)}) } \bigg)^{-1}$.
        
    \IF{$E_{\mathrm{rout}}^{ \mathrm{ISR}} > E_{\mathrm{opt}}^{\mathrm{ISR}}$}
    
    \STATE $E_{\mathrm{opt}}^{\mathrm{ISR}} \leftarrow E_{\mathrm{rout}}^{ \mathrm{ISR}}$, $N_{\mathrm{ISR}}^* \leftarrow N_{\mathrm{ISR}}$, $\theta_{{\mathrm{ISR}},1}^* \leftarrow \theta_{{\mathrm{ISR}},1}$, $\theta_{{\mathrm{ISR}},2}^* \leftarrow \theta_{{\mathrm{ISR}},2}$.
    \ENDIF
    \ENDIF

    \STATE $\theta_{{\mathrm{ISR}},2} \leftarrow \theta_{{\mathrm{ISR}},2} + \frac{1}{N_{\mathrm{in}}} \min \left\{ \theta_{\max}^{(2)}, \Theta-\theta_{{\mathrm{ISR}},1} \right\}$.
    \ENDWHILE
    \STATE $\theta_{{\mathrm{ISR}},1} \leftarrow \theta_{{\mathrm{ISR}},1} + \frac{1}{N_{\mathrm{in}}} \min \left\{ \theta_{\max}^{(1)}, \Theta \right\}$.
    \ENDWHILE
    
    \STATE $N_{\mathrm{ISR}} \leftarrow N_{\mathrm{ISR}} + 1$.
	\ENDWHILE
 
	\STATE \textbf{Output}: Optimal values of decision variables $\left\{ N_{\mathrm{ISR}}^*, \, \theta_{{\mathrm{ISR}},1}^*, \, \theta_{{\mathrm{ISR}},2}^* \right\}$ in ISR.
	\end{algorithmic}
\end{algorithm}

In Algorithm~\ref{alg1}, $\theta_{\max}^{(Q)}$ in step (2) denotes the maximum central angle that can maintain stable communication. $N_{\max}^{\mathrm{STR}}$ in step (3) is a preset maximum number of hops. In step (5), $\theta_{{\mathrm{STR}},1} \leq \frac{2\Theta}{N_{\mathrm{STR}}}$ ensures $\theta_{{\mathrm{STR}},2} = \frac{2\Theta}{N_{\mathrm{STR}}} - \theta_{{\mathrm{STR}},1} \geq 0$. Considering that the Euclidean distance of each hop has been elongated by a factor of $\alpha(\theta)$, steps (7)-(8) correspondingly increase $\theta_{{\mathrm{STR}},Q_1}$ to $\widetilde{\theta}_{{\mathrm{STR}},Q_1}^{(Q_2)}$. In step (9), $\varepsilon$ is a small value used to decrease the probability of links exceeding the maximum communicable distance. On the right-hand side of the arrow in step (10), the four terms represent the reciprocals of the average energy efficiency for the first hop, last hop, even-numbered middle hops, and odd-numbered middle hops. $N_{\mathrm{in}}$ in step (15) is the number of iterations in the inner loop. Finally, since $N_{\mathrm{STR}}$ is an even integer, it increments by $2$ at each iteration in step (17).

Since Algorithm~\ref{alg1} and Algorithm~\ref{alg2} follow comparable approaches and execution steps.  Algorithm~\ref{alg2} involves three decision variables, whereas Algorithm~\ref{alg1} only includes two. Consequently, Algorithm~\ref{alg2} requires an extra loop compared to Algorithm~\ref{alg1}. Based on Algorithsm~\ref{alg1} and \ref{alg2}, we propose in Algorithm~\ref{alg3} a relay subset selection algorithm, which can be applied to both STR and ISR. 
\begin{algorithm}[!ht] 
    \caption{Relay Subset Selection Algorithm}
	\label{alg3} 

	\begin{algorithmic} [1]
    \STATE Search for the optimal values of decision variables using Algorithm~\ref{alg1} or Algorithm~\ref{alg2}. 

    \STATE Based on the decision valuables and Proposition~\ref{propsition1} or Proposition~\ref{propsition2}, identify the optimal relay positions.

    \STATE Search for the relay devices nearest to the ideal relay positions to form the route.

    \STATE If the distance of a hop within the route exceeds the maximum distance that can maintain stable communication, then add one (for ISR) or two (for STR) relays to that specific hop. The relay selection is based on the minimum deviation angle strategy proposed in \cite{wang2022stochastic}.
 
	\STATE \textbf{Output}: Relay positions of the route.
	\end{algorithmic}
\end{algorithm}	

As for the computational complexity of Algorithm~\ref{alg3}, it primarily lies in step (1), wherein the most extensive computation is notably the calculation of the distance scale factor, which includes a double integral. Only step (8) in Algorithm~\ref{alg1} and step (10) in Algorithm~\ref{alg2} involves the calculation of distance scale factor. Therefore, we consider one execution of either step (8) in Algorithm~\ref{alg1} or step (10) in Algorithm~\ref{alg2} as one unit of computational complexity. As a result, the computational complexity of Algorithm~\ref{alg1}, Algorithm~\ref{alg2}, and Algorithm~\ref{alg3} are $\mathcal{O}(N_{\max}^{\mathrm{STR}} N_{\mathrm{in}}) $, $\mathcal{O}(N_{\max}^{\mathrm{ISR}} N_{\mathrm{in}}^2)$, and $\mathcal{O}(N_{\max}^{\mathrm{STR}} N_{\mathrm{in}} + N_{\max}^{\mathrm{ISR}} N_{\mathrm{in}}^2)$, respectively. In simulation, $N_{\mathrm{in}}$ is set to $20$, since the obtained routing performance is close enough to the optimal value. 

\par
Traditional graph-based routing methods like Dijkstra typically need to simulate the performance of links across the entire LEO satellite constellation to determine the optimal path. Additionally, due to the randomness of the channel fading, at least thousands of routing simulations are required to achieve stable average performance results. Therefore, for a mega-constellation with thousands of satellites, the number of rounds for link performance evaluations is expected to be larger than $10^8$. In contrast, the proposed algorithm requires between $10^3$ and $10^4$ iterations for energy efficiency estimation. From the above analysis, it can be concluded that there are two main reasons why the computational complexity of the routing design in this article is significantly lower than that of current methods. First, when addressing the randomness of channel fading, we use analytical expressions instead of extensive rounds of simulations. Second, for a connected large-scale constellation, the number of links is often much greater than the number of satellites, and modeling these links is computationally expensive. The proposed method avoids modeling the routing links and instead directly identifies the potential optimal relay positions.

\par
Finally, we provide the following remark on the values for the maximum hops,  $N_{\max}^{\mathrm{STR}}$ and $N_{\max}^{\mathrm{ISR}}$. Taking STR as an example, steps (7)-(8) of Algorithm~\ref{alg1} indicate that the value of $\widetilde{\theta}_{{\mathrm{STR}},Q_1}^{(Q_2)}$ is determined by $\theta_{{\mathrm{STR}},1}$ and $\theta_{{\mathrm{STR}},2}$, which, in turn, are closely related to $N_{\mathrm{STR}}$, as can be inferred from steps (5)-(6). Consequently, we limit the values of $N_{\max}^{\mathrm{STR}}$ by setting upper bounds for $\widetilde{\theta}_{{\mathrm{STR}},Q_1}^{(Q_2)}$. 

\begin{remark}\label{remark1}
When the number of hops is fixed as $N_{\max}^{\mathrm{STR}}$ or $N_{\max}^{\mathrm{ISR}}$, executing the inner loop (steps (5)-(16) for Algorithm~\ref{alg1} and steps (5)-(20) for Algorithm~\ref{alg2}) can yield the optimal values for decision variables ($\left\{\theta_{{\mathrm{STR}},1}^*, \, \theta_{{\mathrm{STR}},2}^* \right\}$ for STR and $\left\{\theta_{{\mathrm{ISR}},1}^*, \, \theta_{{\mathrm{ISR}},2}^* \, \theta_{{\mathrm{ISR}},3}^* \right\}$ for ISR). If the inequalities about the increased central angles corresponding to these decision variables are satisfied, then the value of $N_{\max}^{\mathrm{STR}}$ or $N_{\max}^{\mathrm{ISR}}$ is reasonable:
\begin{align}
    \left\{
 	\begin{array}{lll}
    \widetilde{\theta}_{{ \mathrm{STR}},1}^{(2)} < 2\theta_{{ \mathrm{STR}},1} + \theta_{{ \mathrm{STR}},2}, \ \widetilde{\theta}_{{ \mathrm{STR}},2}^{(2)} < \theta_{{ \mathrm{STR}},1} + 2\theta_{{ \mathrm{STR}},2}, \\ \ \ \ \ \ \ \ \ \ \ \ \ \ \ \ \ \ \ \ \ \ \ \ \ \ \ \ \ \ \ \ \ \ \ \ \ \ \ \ \ \ \ \ \ \ \ \ \ \ \ {\mathrm{for \ }} N_{\max}^{\mathrm{STR}}, \\
    \widetilde{\theta}_{{ \mathrm{ISR}},Q}^{(2)} < \theta_{{ \mathrm{ISR}},Q} + \theta_{{ \mathrm{ISR}},3}, \, Q=\{1,2\}, \ \widetilde{\theta}_{{ \mathrm{ISR}},3}^{(3)} < 2\theta_{{ \mathrm{ISR}},3}, \\ \ \ \ \ \ \ \ \ \ \ \ \ \ \ \ \ \ \ \ \ \ \ \ \ \ \ \ \ \ \ \ \ \ \ \ \ \ \ \ \ \ \ \ \ \ \ \ \ \ \ {\mathrm{for \ }} N_{\max}^{\mathrm{ISR}}.
	\end{array}
	\right.
\end{align}
\end{remark}
As the number of hops increases, the distance of each hop becomes shorter, and the closest relay to adjacent ideal relay positions could be the same one. In this case, it would result in repeated signal transmission through the same relay device, leading to the wastage of energy. When the inequalities in the remark are satisfied, the probability of this event occurring is relatively low. Remark~\ref{remark1} only provides a necessary but not sufficient condition for the reasonableness of the maximum number of hops. This is understandable because when there are few relays, there might be no relays within the maximum distance range. This implies that the routing might be unavailable regardless of how the maximum number of hops is set.

\subsection{Performance Evaluation}
This subsection analyzes the relay availability and routing energy efficiency of Algorithm~\ref{alg3}. 
First, we define the routing availability probability, which is used to measure relay availability, as follows.

\begin{definition}[Routing Availability Probability]
    Given that a route is obtained by  Algorithm~\ref{alg3}, the routing availability probability is defined as the probability that the distance for each hop in the route is less than the maximum distances that can maintain stable communication. 
\end{definition}

To start with, we introduce the single-hop availability probability provided as Corollary 2 in \cite{wang2022conditional}. Note that this corollary is specific to the scenario where two devices are ground GWs, and the relay is a satellite. Therefore, the existing results need to be expanded.

\begin{lemma}\label{lemma3}
Given the central angle between two communication devices $\theta_c$, the probability of the existence of at least one relay between these two devices is  
\begin{equation}\label{lemma3-1}
\begin{split}
    & P_Q^A(\theta_1, \theta_2, \theta_c) = 1 - \\
    & \left( 1 - \frac{ S_Q(\theta_1,\psi_1) +  S_Q(\theta_2,\psi_2) }{4 \pi R_Q^2} \right)^{N_Q}, \ Q \in \{g,s\},
\end{split}
\end{equation}
where $\theta_1$ and $\theta_2$ represent the maximum central angles within which the relay can maintain reliable communication with two devices, $Q$ is assigned with labels $s$ and $g$ when the relay device is a satellite and a GW, respectively; the function 
$S_Q(\theta,\psi)$ is given as
\begin{equation}
\begin{split}
    & S_Q(\theta,\psi) = \int_{R_Q \cos\theta \tan(\theta-\psi) }^{ R_Q \sin\theta } 2 R_Q \\
    & \times \arcsin\left( \frac{1}{R_Q} \sqrt{R_Q^2 (\sin\theta)^2 - l^2 } \right) {\mathrm{d}}l.
\end{split}
\end{equation}
whereby denoting $x = \cos\theta_1$ and $y = \cos\theta_2$, $\psi_1$ and $\psi_2$ are given as:
\begin{align}\label{theo1-6}
    \left\{
 	\begin{array}{lll}
    \psi_1 = \theta_1 - \frac{1}{x-y} \left( x \, \theta_c - \sqrt{2x^2 - 4xy + 2y^2 + xy\theta_c^2} \right), \\
    \psi_2 = \theta_2 - \frac{1}{x-y} \left( y \, \theta_c - \sqrt{2x^2 - 4xy + 2y^2 + xy\theta_c^2} \right).
	\end{array}
	\right.
\end{align}
\end{lemma}
\begin{IEEEproof}
The derivation follows a similar procedure to \cite{wang2022conditional}, therefore omitted here. 
\end{IEEEproof}

\par
Note that for the sake of convenience in the statement, we assumed that $R_{\oplus}$ and $R_g$ have the same meanings in Lemma~\ref{lemma3}. Next, the set of decision variables denoted as $\{N_{\mathrm{STR}}, \theta_{{\mathrm{STR}},1}, \theta_{{\mathrm{STR}},2}\}$ is an STR strategy obtained through Algorithm~\ref{alg3}, consisting of $N_{\mathrm{STR}}$ hops, and the central angles for the ideal relay positions at odd and even hops are  $\theta_{{\mathrm{STR}},1}$ and $\theta_{{\mathrm{STR}},2}$, respectively.

\begin{theorem}\label{theorem1}
For an STR with the set of decision variables $\{N_{\mathrm{STR}}, \theta_{{\mathrm{STR}},1}, \theta_{{\mathrm{STR}},2}\}$, the routing availability probability can be estimated as (\ref{theo1-1}) at the top of next page, 
\begin{table*}
\begin{equation}\label{theo1-1}
\begin{split}
    & P_{\mathrm{STR}}^A = P_g^A \left( \theta_{{\mathrm{STR}},1}^{\max}, \theta_{{\mathrm{STR}},2}^{\max}, \theta_{{\mathrm{STR}},1} +  \widetilde{\alpha}_g (\theta_{{\mathrm{STR}},2}) \, \theta_{{\mathrm{STR}},2} \right) \times P_g^A \left( \theta_{{\mathrm{STR}},1}^{\max}, \theta_{{\mathrm{STR}},2}^{\max}, \theta_{{\mathrm{STR}},2} +  \widetilde{\alpha}_g (\theta_{{\mathrm{STR}},1}) \, \theta_{{\mathrm{STR}},1} \right) \\
    & \times \left( P_g^A \left( \theta_{{\mathrm{STR}},1}^{\max}, \theta_{{\mathrm{STR}},2}^{\max}, \widetilde{\alpha}_g (\theta_{{\mathrm{STR}},1}) \, \theta_{{\mathrm{STR}},1} +  \widetilde{\alpha}_g (\theta_{{\mathrm{STR}},2}) \, \theta_{{\mathrm{STR}},2} \right) \right)^{ \max \left\{0, \frac{N_{\mathrm{STR}}}{2} - 1 \right\} } \\
    & \times \left( P_s^A \left( \theta_{{\mathrm{STR}},1}^{\max}, \theta_{{\mathrm{STR}},2}^{\max}, \widetilde{\alpha}_s (\theta_{{\mathrm{STR}},1}) \, \theta_{{\mathrm{STR}},1} +  \widetilde{\alpha}_s (\theta_{{\mathrm{STR}},2}) \, \theta_{{\mathrm{STR}},2} \right) \right)^{ \max \left\{0, \frac{N_{\mathrm{STR}}}{2} - 2 \right\} },
\end{split}
\end{equation}
\hrule
\end{table*}
where $P_Q^A(\theta_1, \theta_2, \theta_c)$ is defined in Lemma~\ref{lemma3}, $\theta_{{\mathrm{STR}},Q}^{\max} =  \arccos\left( \frac{R_s^2 + R_{\oplus}^2 - \big( l_{\max}^{(Q)} \big)^2 }{ 2 R_s R_{\oplus} } \right), \ Q=\{1,2\}$,  $\widetilde{\alpha}_g (\theta)$ and $\widetilde{\alpha}_s (\theta)$ are given as follows:
\begin{align}\label{theo1-2}
    \left\{
 	\begin{array}{lll}
    \widetilde{\alpha}_g(\theta) = \frac{1}{\theta} \times \\ \arccos\left( \frac{2R_s R_{\oplus}\left( \alpha^{(2)}(\theta) \right)^2 - \left( \left( \alpha^{(2)}(\theta) \right)^2 - \left( \alpha^{(1)}(\theta) \right)^2 \right)(R_s^2 + R_{\oplus}^2 ) }{2R_s R_{\oplus} \left( \alpha^{(1)}(\theta) \right)^2 } \right), \\
    \widetilde{\alpha}_s(\theta) = \frac{1}{\theta} \times \\ \arccos\left( \frac{2R_s R_{\oplus}\left( \alpha^{(1)}(\theta) \right)^2 - \left( \left( \alpha^{(1)}(\theta) \right)^2 - 1 \right)(R_s^2 + R_{\oplus}^2 ) }{2R_s R_{\oplus} } \right),
	\end{array}
	\right.
\end{align}
where $\alpha^{(1)}(\theta)$ and $\alpha^{(2)}(\theta)$ are defined in Lemma~\ref{lemma2}.
\end{theorem}
\begin{IEEEproof}
    See Appendix~\ref{app:theorem1}.
\end{IEEEproof}

Similarly, an ISR strategy can be denoted as the set $\{N_{\mathrm{ISR}}, \theta_{{\mathrm{ISR}},1}, \theta_{{\mathrm{ISR}},2}, \theta_{{\mathrm{ISR}},3}\}$, where $\theta_{{\mathrm{ISR}},1}$, $\theta_{{\mathrm{ISR}},2}$ and  $\theta_{{\mathrm{ISR}},3}$ represent the central angles of the first, last and middle hops, respectively.

\begin{theorem}\label{theorem2}
For an ISR with the set of decision variables $\{N_{\mathrm{ISR}}, \theta_{{\mathrm{ISR}},1}, \theta_{{\mathrm{ISR}},2}, \theta_{{\mathrm{ISR}},3} \}$, the routing availability probability can be estimated as,
\begin{sequation}
\begin{split}
    & P_{\mathrm{ISR}}^A = P_s^A \left( \theta_{{\mathrm{ISR}},1}^{\max}, \theta_{{\mathrm{ISR}},3}^{\max}, \theta_{{\mathrm{STR}},1} +  \widetilde{\alpha}_s (\theta_{{\mathrm{ISR}},3}) \, \theta_{{\mathrm{ISR}},3} \right) \\ 
    & \times P_s^A \left( \theta_{{\mathrm{ISR}},2}^{\max}, \theta_{{\mathrm{ISR}},3}^{\max}, \theta_{{\mathrm{ISR}},2} +  \widetilde{\alpha}_s (\theta_{{\mathrm{ISR}},3}) \, \theta_{{\mathrm{ISR}},3} \right) \\
    & \times \left( P_s^A \left( \theta_{{\mathrm{ISR}},3}^{\max}, \theta_{{\mathrm{ISR}},3}^{\max}, 2 \widetilde{\alpha}_s (\theta_{{\mathrm{ISR}},3}) \, \theta_{{\mathrm{ISR}},3} \right) \right)^{N_{\mathrm{ISR}} - 2},
\end{split}
\end{sequation}where $\widetilde{\alpha}_s (\theta)$ is defined in (\ref{theo1-2}). $\theta_{{\mathrm{ISR}},1}^{\max} = \theta_{{\mathrm{STR}},1}^{\max}$, $\theta_{{\mathrm{ISR}},2}^{\max} = \theta_{{\mathrm{STR}},2}^{\max}$, and $\theta_{{\mathrm{ISR}},3}^{\max} = 2 \arcsin \left( {l_{\max}^{(3)}}/{2R_s} \right)$. 
\end{theorem}
\begin{IEEEproof}
    The proof of Theorem~\ref{theorem2} is similar to that of Theorem~\ref{theorem1}, therefore omitted here.
\end{IEEEproof}

\par
If the routing availability probability in Theorem~\ref{theorem1} or Theorem~\ref{theorem2} is low, the value of $\varepsilon$ should be appropriately reduced. Recall that $\varepsilon$ is applied in step (9), Algorithm~\ref{alg1} and step (11), Algorithm~\ref{alg2}. Next, the analytical expressions of contact angle distributions, which serve as important lemmas for routing energy efficiency, are provided.

\begin{lemma}\label{lemma4}
Given that the central angle of a hop is $\phi$. For the first and last hops in STR and ISR, the central angle distribution is
\begin{equation}
\begin{split}
    & f_{\theta_c^{(1)}} \left( \theta_c \, | \, \phi \right) = \int_0^{2\pi} \frac{N_s \sin\theta_c }{4 \pi} \\
    & \times \left( \frac{1}{2} + \frac{ R_s^2 + R_{\oplus}^2 -  ( d( R_s,\theta_c,\varphi; R_{\oplus},\phi,0))^2 }{4 R_s R_{\oplus}} \right)^{N_s-1} \mathrm{d}\varphi,
\end{split}
\end{equation}
where $d( R_s,\theta_c,\varphi; R_{\oplus},\phi,0)$ is defined in (\ref{d_12}). For middle hops in STR, an approximation for the central angle distribution is
\begin{equation}\label{lemma4-2}
\begin{split}
    f_{\theta_c^{(2)}} \left( \theta_c \, | \, \phi \right) =  \frac{f_{\theta_c^{(1)}} \left( \arccos\Xi  \, | \, \phi \right) \sin\theta_c}{\left( \alpha^{(1)}(\phi) \right)^2 \sqrt{1-\Xi^2}},
\end{split}
\end{equation}
where $\Xi$ in (\ref{lemma4-2}) is defined as
\begin{sequation}
    \Xi = \frac{R_s^2 + R_{\oplus}^2 - \left( \alpha^{(1)}(\phi) \right)^{-2} \left( R_s^2 + R_{\oplus}^2 - 2R_s R_{\oplus} \cos\theta_c \right)}{2R_s R_{\oplus}}.
\end{sequation}
For middle hops in ISR, an approximation for the central angle distribution is
\begin{equation}\label{lemma4-4}
\begin{split}
    f_{\theta_c^{(3)}} \left( \theta_c \, | \, \phi \right) & = f_{\widetilde{\theta}_c^s} \left(2 \arcsin \left( \frac{\sin(\theta_c/2)}{ \sqrt{\alpha^{(3)}\left( \phi \right)} } \right) \right) \\
    & \times \frac{\cos(\theta_c/2)}{\sqrt{\alpha^{(3)}\left( \phi \right)  - \sin^2(\theta_c/2)}},
\end{split}
\end{equation}
where $f_{\widetilde{\theta}_c^s} \left( \theta \right)$ in (\ref{lemma4-4}) is defined as,
\begin{sequation}\label{lemma4-5}
\begin{split}
    &f_{\widetilde{\theta}_c^s} \left( \theta \right) = \int_0^{2\pi} \frac{N_s \sin\theta}{2^{N_s+1} \pi} \\
    & \times \left( 1 + \cos\left( 2\arcsin \left( \frac{d(R_s,\theta,\psi ; R_s,\frac{\Theta}{N_l},0)} {2R_s} \right) \right) \right)^{N_s-1} \mathrm{d}\psi.
\end{split}
\end{sequation}
\end{lemma}
\begin{IEEEproof}
    See Appendix~\ref{app:lemma4}.
\end{IEEEproof}

\par
Based on the contact angle distributions, the analytical results of routing energy efficiency can be derived from the following theorems.

\begin{theorem}\label{theorem3}
For an ISR with the set of decision variables $\{N_{\mathrm{ISR}}, \theta_{{\mathrm{ISR}},1}, \theta_{{\mathrm{ISR}},2}, \theta_{{\mathrm{ISR}},3} \}$, the routing energy efficiency can be approximated as
\begin{sequation}
\begin{split}
    & E_{\mathrm{rout}}^{\mathrm{ISR}} = \bigg( \int_0^{\pi} f_{\theta_c^{(1)}}(\psi \, | \,\theta_{{\mathrm{ISR}},1}) \overline{E}_{\mathrm{hop}}^{(1)}(\theta_{{\mathrm{ISR}},1}) \mathrm{d}\psi \\
    & + \int_0^{\pi} f_{\theta_c^{(1)}}(\psi \, | \,\theta_{{\mathrm{ISR}},2}) \overline{E}_{\mathrm{hop}}^{(2)}(\theta_{{\mathrm{ISR}},2}) \mathrm{d}\psi \\
    & + \left( N_{\mathrm{ISR}} - 2 \right) \int_0^{\pi} f_{\theta_c^{(3)}}(\psi \, | \,\theta_{{\mathrm{ISR}},3}) \overline{E}_{\mathrm{hop}}^{(3)}(\theta_{{\mathrm{ISR}},3}) \mathrm{d}\psi \bigg)^{-1},
\end{split}
\end{sequation}
where $f_{\theta_c^{(1)}} \left( \theta_c \, | \, \phi \right)$ and $f_{\theta_c^{(3)}} \left( \theta_c \, | \, \phi \right)$ are defined in Lemma~\ref{lemma4}, and $\overline{E}_{\mathrm{hop}}^{(Q)}(\theta), Q=\{1,2,3\}$, are defined in Lemma~\ref{lemma1}.
\end{theorem}
\begin{IEEEproof}
    See Appendix~\ref{app:theorem3}.
\end{IEEEproof}
 
Similarly, the analytical expression for the routing energy efficiency of STR can be obtained as follows.

\begin{table*}[]
\centering
\caption{Simulation Parameters \cite{talgat2020stochastic,ata2022performance}.}
\label{table1}
\resizebox{\linewidth}{!}{ 
\renewcommand{\arraystretch}{1.1}
\begin{tabular}{|c|c|c||c|c|c|}
\hline
Notation     & Meaning                         & Default Value     & Notation   & Meaning    & Default Value         \\ \hline \hline
$N_g,N_s$ & Number of GWs, satellites     & 1000 &  $\beta$ & Price ratio factor  & 5 \\ \hline
$R_{\oplus},R_s$ & Radius of the Earth, satellites     & $6371,7371$~km &  $\sigma_s^2, \sigma_g^2$ & Noise power  & $10^{-10}$~mW \\ \hline
$\Omega,b_0,n_0$  &  Parameters of the SR fading  &  $1.29,0.158,19.4$  & $\zeta_{\mathrm{ST}}$, $\zeta_{\mathrm{SS}}$ & Rain attenuation & $-2$~dB, $0$~dB \\ \hline
$\eta_s$, $A_0$ & Parameters of the pointing error                  & $1.00526,3.2120$ %0.01979 
& $G_{\mathrm{ST}}, G_{\mathrm{SS}}$ & Antenna gain   & $41.7$~dBi           \\ \hline
$\varsigma$  & Variance of Rayleigh distribution & $15$~mrad & $\lambda_{\mathrm{ST}}, \lambda_{\mathrm{SS}}$  & Wavelength   & $1550$~nm     \\ \hline
$\Theta$     & {\scriptsize{Central angle between the transmitter and receiver}}    & $\pi$    & $B_{\mathrm{ST}}, B_{\mathrm{SS}}$ & Bandwidth    & $20$~MHz     \\ \hline
$l_{\max}^{(1)},l_{\max}^{(2)},l_{\max}^{(3)}$   & Maximum distance of communication   & $3000$~km  & $\rho_t^{(1)},\rho_t^{(2)},\rho_t^{(3)}$  & Transmission power & $15$~dBW \\ \hline
\end{tabular}
}
\end{table*}

\begin{theorem}\label{theorem4}
For an STR with the set of decision variables $\{N_{\mathrm{STR}}, \theta_{{\mathrm{STR}},1}, \theta_{{\mathrm{STR}},2}\}$, the routing energy efficiency can be approximately estimated as
\begin{sequation}
\begin{split}
    & E_{\mathrm{rout}}^{\mathrm{STR}} = \bigg( \int_0^{\pi} f_{\theta_c^{(1)}}(\psi \, | \, \theta_{{\mathrm{STR}},1}) \overline{E}_{\mathrm{hop}}^{(1)}(\theta_{{\mathrm{STR}},1}) \mathrm{d}\psi \\
    & + \left( \frac{N_{\mathrm{STR}}}{2} - 1 \right) \int_0^{\pi} f_{\theta_c^{(2)}}(\psi \, | \,\theta_{{\mathrm{STR}},1}) \overline{E}_{\mathrm{hop}}^{(1)}(\theta_{{\mathrm{STR}},1}) \mathrm{d}\psi \\ 
    & + \left( \frac{N_{\mathrm{STR}}}{2} - 1 \right) \int_0^{\pi} f_{\theta_c^{(2)}}(\psi \, | \,\theta_{{\mathrm{STR}},2}) \overline{E}_{\mathrm{hop}}^{(2)}(\theta_{{\mathrm{STR}},2}) \mathrm{d}\psi \\
    & + \int_0^{\pi} f_{\theta_c^{(1)}}(\psi \, | \, \theta_{{\mathrm{STR}},2}) \overline{E}_{\mathrm{hop}}^{(2)}(\theta_{{\mathrm{STR}},2}) \mathrm{d}\psi \bigg)^{-1},
\end{split}
\end{sequation}
where $f_{\theta_c^{(1)}} \left( \theta_c \, | \, \phi \right)$ and $f_{\theta_c^{(2)}} \left( \theta_c \, | \, \phi \right)$ are defined in Lemma~\ref{lemma4}, $\overline{E}_{\mathrm{hop}}^{(1)}(\theta)$ and $\overline{E}_{\mathrm{hop}}^{(2)}(\theta)$ are defined in Lemma~\ref{lemma1}.
\end{theorem}
\begin{IEEEproof}
    The proof of Theorem~\ref{theorem4} is similar to that of Theorem~\ref{theorem3}, therefore omitted here.
\end{IEEEproof}

\section{Numerical Results}
This section provides the numerical results of routing availability probability and energy efficiency. Unless otherwise specified, the parameters are set to their default values in Table~\ref{table1}. 
It is worth mentioning that the maximum distance of communication is fixed at $3000$~km based on references \cite{sag2018modelling} and \cite{wang2022stochastic}. However, the single-hop distance rarely reaches this upper limit when the energy efficiency is maximized. 

\begin{figure}[ht]
\centering
\includegraphics[width=0.9\linewidth]{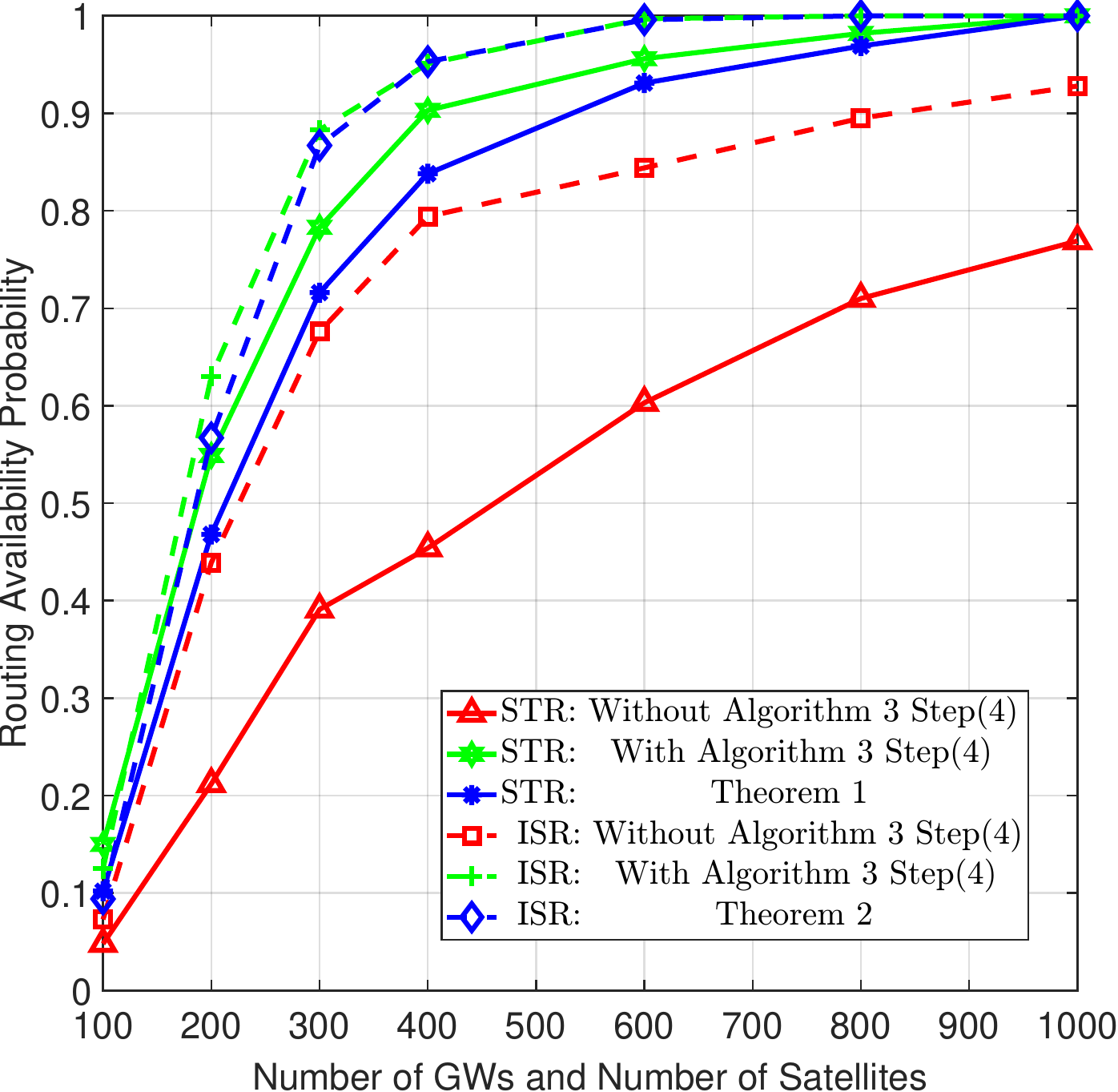}
\caption{Routing availability probability with different numbers of GWs and satellites.}
\label{fig2}
\end{figure}

\subsection{Routing Availiability Probability}
Fig.~\ref{fig2} compares the routing availability probability of ISR and STR. The label "Without Algorithm~\ref{alg3} Step(4)" refers to executing only the first three steps of Algorithm~\ref{alg3}. The labels "Theorem~\ref{theorem1}" and "Theorem~\ref{theorem2}" refer to estimating the availability probability with low complexity through analytical methods, without running Algorithm~\ref{alg3}. Denote the $x$-axis "Number of GWs and Number of Satellites" as $N_{\mathrm{total}}$. For a fair comparison, we assume $N_g = 0$  and $N_s = N_{\mathrm{total}}$ for ISL, $N_g = N_s = \frac{1}{2} N_{\mathrm{total}}$ for STR.

\par
As shown in the figure, step(4) significantly improves the routing availability, especially for STR. With the same number of satellites, ISR has a higher availability probability than STR. The analytical results provided by Theorem~\ref{theorem1} and Theorem~\ref{theorem2} offer a relatively close lower bound for the availability probability when the algorithm is completely executed. When the number of satellites in the constellation exceeds $600$ and $1000$ respectively, the availability probability for ISR and STR can reach $100\%$. It is worth noting that in the simulation if any hop in the routing is unavailable, the energy efficiency of the route is considered to be $0$.

\begin{figure}[ht]
\centering
\includegraphics[width=0.9\linewidth]{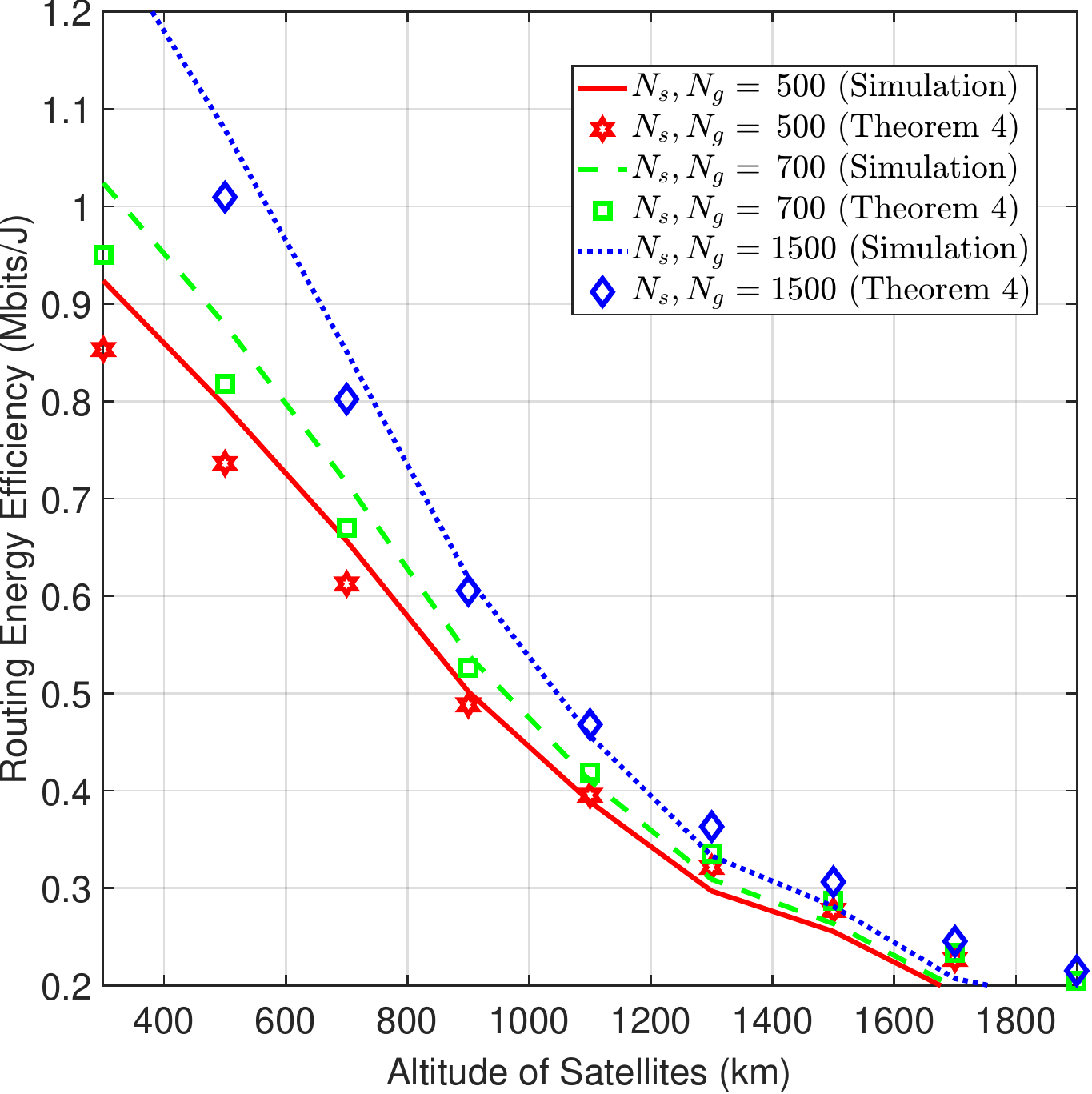}
\caption{Routing energy efficiency of STR with different constellation configurations.}
\label{fig3}
\end{figure}

\subsection{Routing Energy Efficiency}
\begin{figure}[ht]
\centering
\includegraphics[width=0.9\linewidth]{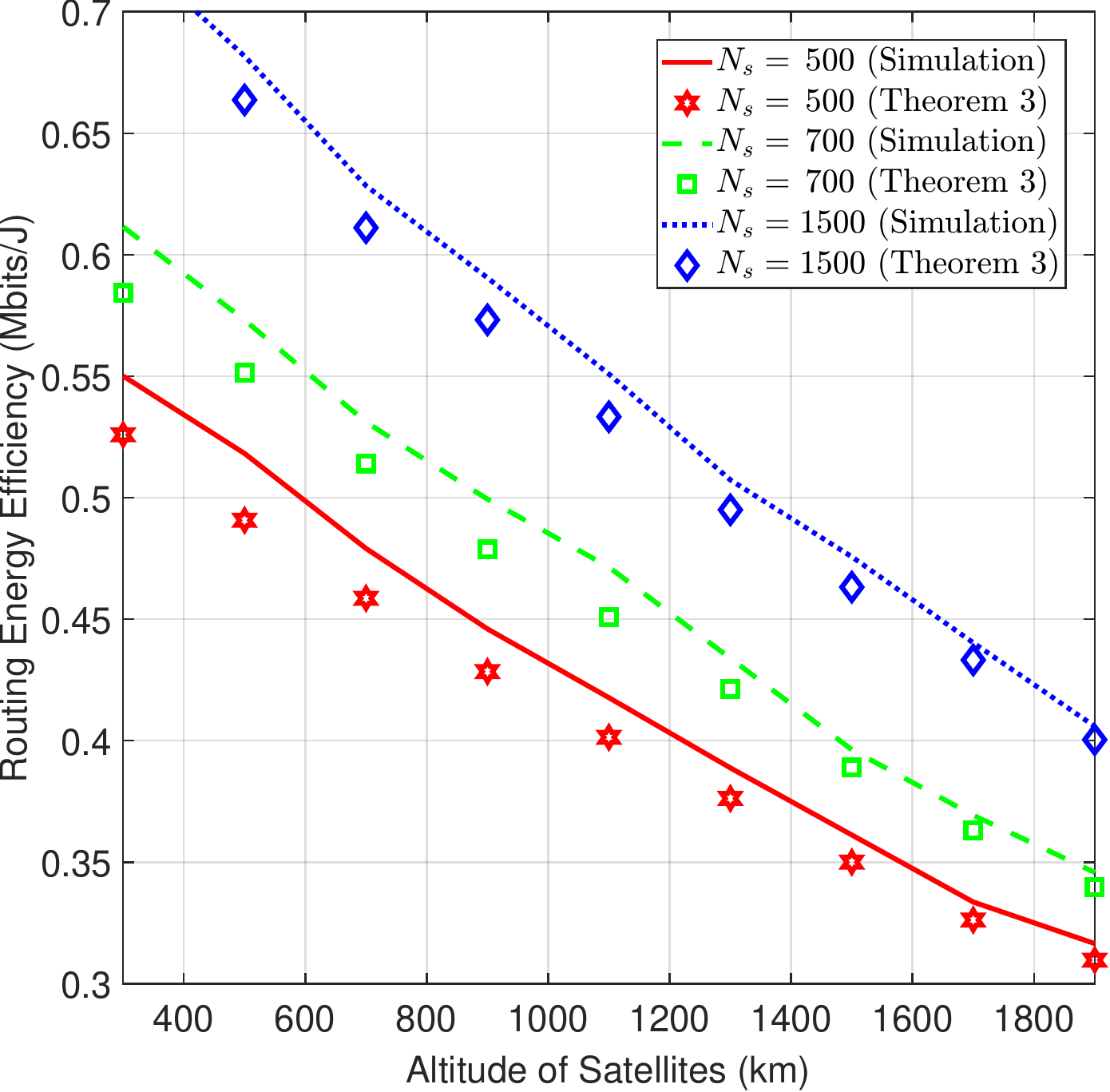}
\caption{Routing energy efficiency of ISR with different constellation configurations.}
\label{fig4}
\end{figure}

This subsection first explains how Fig.~\ref{fig3} and Fig.~\ref{fig4} verify the accuracy of the approximation given by the theorems. Since outliers in small-scale fading values in simulations can lead to extreme results in the reciprocal of single-hop energy efficiency,  outliers are removed based on the three times standard deviation criterion. Then, we define the relative error as the ratio of the absolute difference between the energy efficiency obtained from the simulation and analytical results to the energy efficiency estimated by the simulation. The relative error between the analytical results provided in Theorem~\ref{theorem4} and the routing energy efficiency of ISR obtained by simulation is $1.54\%$, The relative error between that of Theorem~\ref{theorem3} and the simulation results in STR is $3.41\%$. Due to the acceptable matching between analytical results and simulations, we only show analytical results in the following part of this section.

\par
As shown in Fig.~\ref{fig3} and Fig.~\ref{fig4}, as the number of satellites (and GWs) increases, both STR and ISR show improved routing energy efficiency. A decrease in constellation altitude can have the same effect. The close spacing of the three curves in Fig.~\ref{fig3} indicates that the constellation altitude has a greater impact on energy efficiency compared to the number of satellites for STR. At the same altitude and with the same total number of devices, although the availability performance of STR is worse than that of ISR, STR has a significant advantage in routing energy efficiency.

\begin{figure}[ht]
\centering
\includegraphics[width=0.9\linewidth]{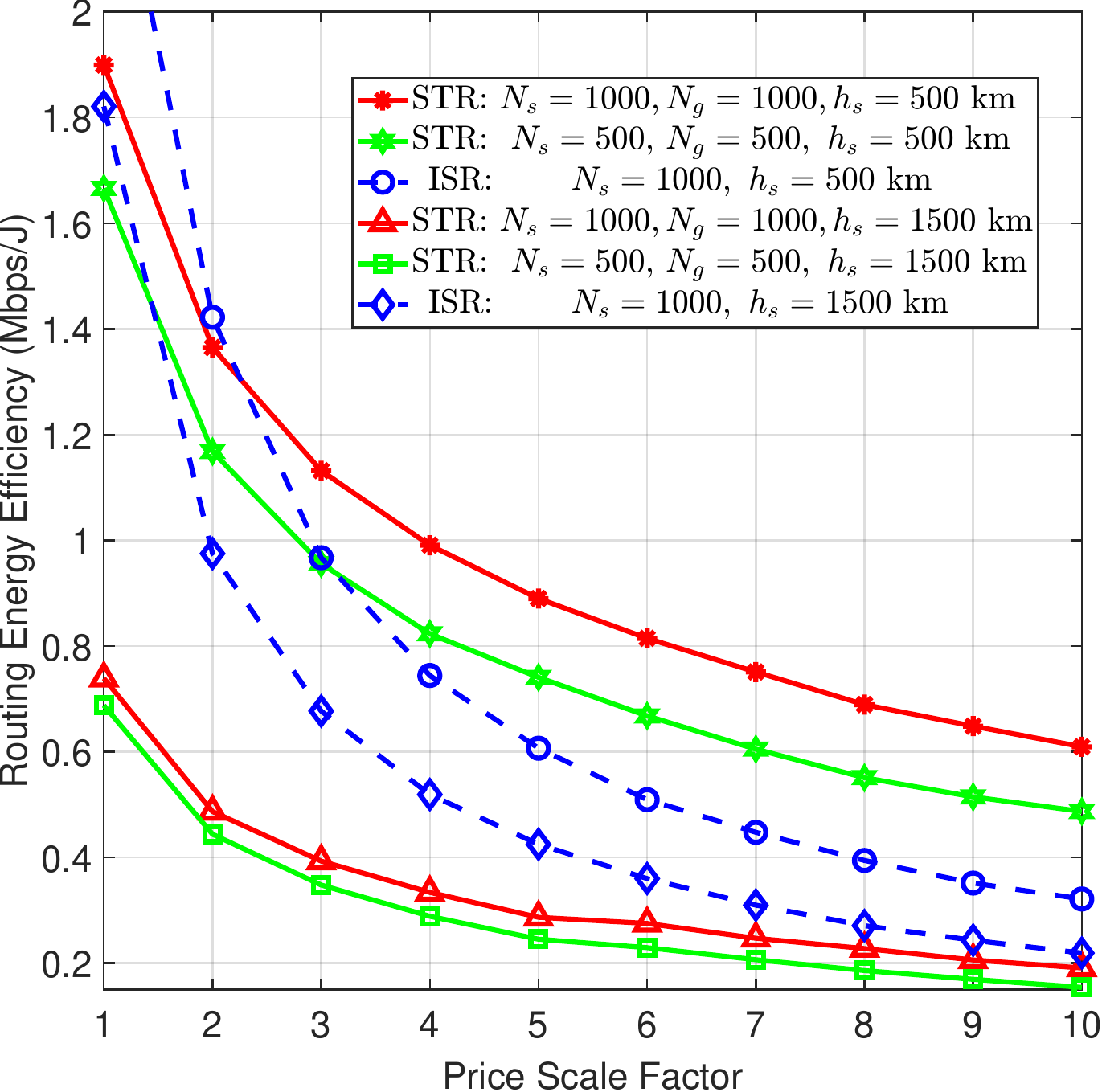}
\caption{Routing energy efficiency with different price ratio factors.}
\label{fig5}
\end{figure}

Fig.~\ref{fig5} shows the impact of the price ratio factor on routing energy efficiency. Recall that the price ratio factor is the ratio between the price of energy in space and that on the ground. When the constellation's altitude is $h_s=1500$~km, ISR consistently outperforms STR in terms of energy efficiency. At $h_s=500$~km, STR exhibits higher energy efficiency than ISR with the same number of satellites when $\beta>2.4$; meanwhile, with the same number of devices, STR demonstrates higher energy efficiency than ISR when $\beta>3$.

\subsection{Strategies Comparison}
\begin{figure}[ht]
\centering
\includegraphics[width=0.9\linewidth]{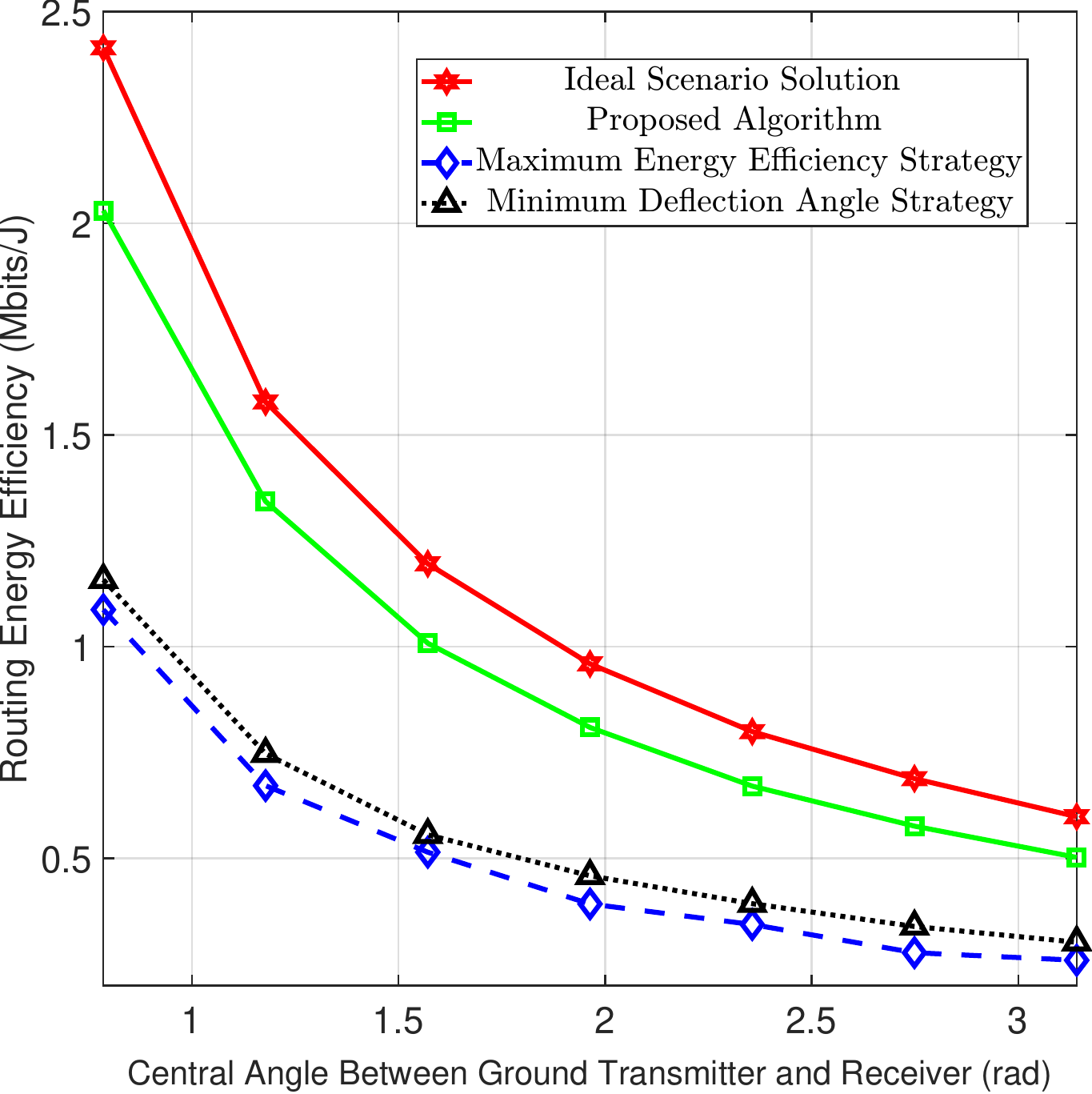}
\caption{Strategies comparison for STR.}
\label{fig6}
\end{figure}

\begin{figure}[ht]
\centering
\includegraphics[width=0.9\linewidth]{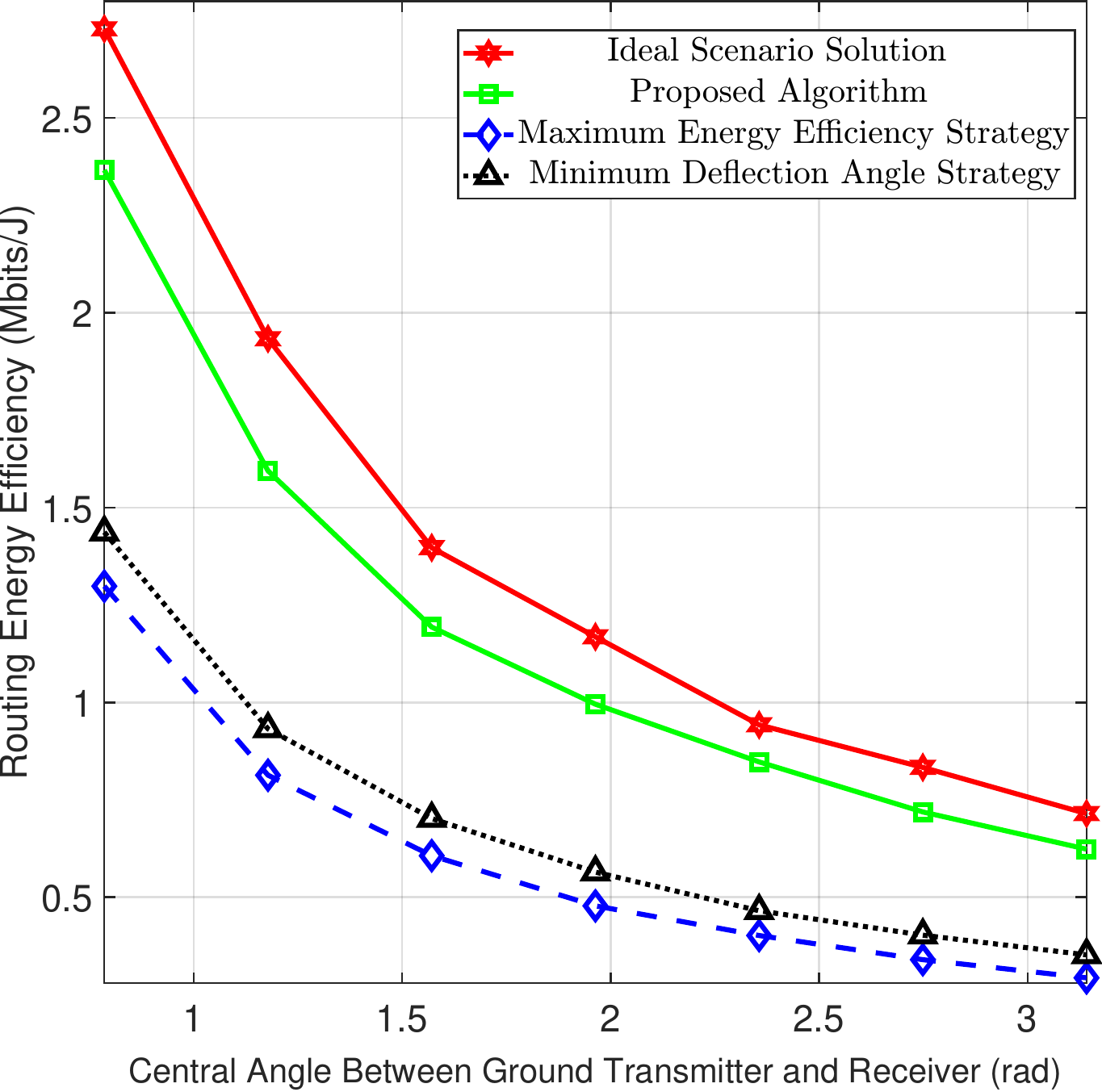}
\caption{Strategies comparison for ISR.}
\label{fig7}
\end{figure}

This subsection compares the proposed algorithm in this article with other recently proposed SG-based routing strategies. {\color{black}Since the performance metrics of the proposed method can be estimated by analytical expressions, we also select algorithms with analyzable performance for comparison.} The labels of Fig.~\ref{fig6} and Fig.~\ref{fig7} are explained as follows.
\begin{itemize}
    \item Ideal Scenario Solution: The method proposed in Proposition~\ref{proposition3}. Since it has been proven to achieve maximum energy efficiency in an ideal scenario, it can be used as an upper bound for reference. However, the assumption that satellites are available at every location is ideal, making this upper bound unattainable in practice.
    \item Maximum Energy Efficiency Strategy: Referring to \cite{lou2023coverage},  each device searches all devices within its communication range and selects the one with the highest single-hop energy efficiency towards the ground receiver as the next hop. Hence, it is also referred to as a greedy algorithm.
    \item Minimum Deflection Angle Strategy: Referring to \cite{wang2022stochastic}, each relay selects the device within the communication range that deviates the least from the shortest inferior arc as the next hop. 
\end{itemize}

Fig.~\ref{fig6} and Fig.~\ref{fig7} indicate that the proposed strategy for STR or ISR has a small gap from the upper bound provided by the ideal scenario solutions. As a result, the estimated routing energy efficiencies in the theorems can still provide tight lower bounds for these potentially better-performing methods. Finally, our proposed algorithm exhibits a significant advantage when compared with existing proposed SG-based routing strategies. One possible reason for this result is that the proposed algorithm designs the routing strategy from a global perspective, whereas the comparison methods rely solely on local topological information.

\section{Conclusion and Future Work}
In this article, we developed an analytical framework to assess the link qualities of satellite-terrestrial and inter-satellite connections, based on which we proposed algorithms for maximizing the routing energy efficiency of STR and ISR. These algorithms demonstrated significant advantages in energy efficiency over existing SG-based routing strategies and could approach the ideal energy efficiency upper bound. 
We found that the comparative energy efficiency between STR and ISR was dependent on the ratio of satellite to ground energy prices. 
Furthermore, we provided several theoretical results to estimate the satellite availability probability and energy efficiency of these algorithms, which were validated through numerical results. 
Our numerical findings also revealed that communication systems with a larger number of devices and lower constellation altitudes achieved higher energy efficiency.

\par
{\color{black}
A potential direction for future work is a more comprehensive comparison between STR and ISR. In this paper, the discussion of both approaches is limited to routing performance at the communication layer, but other factors also deserve attention. For example, due to political reasons, deploying ground GWs globally requires collaboration with local enterprises, operators, and even governments, which introduces additional costs for STR. Additionally, previous studies in spherical SG generally focused more on geometric topology while insufficient attention was given to detailed channel modeling. Similarly, this paper emphasizes routing design based on the spatial distribution of satellites, which led us to adopt a simplified channel model for ease of analysis. In \cite{lin2022refracting,he2022multiobjective}, the authors designed phased arrays to exploit both spatial multiplexing and power control, effectively accelerating networking speed and achieving more efficient transmission. Considering the significance of factors like antenna arrays and the current lack of research in this area within the spherical SG field, advancing the integration of the stochastic channel analytical framework with the SG analytical framework represents another important research direction.
}

\appendices
\section{Proof of Proposition~\ref{propsition1}}\label{app:propsition1}
Denote the $i^{th}$ hop's central angle between communication devices as $\psi_i$. To ensure the endpoint is reachable, inequality $\sum_{i=1}^{N_{\mathrm{STR}}} \psi_i \geq \Theta$ needs to be satisfied. The larger $\psi_i$ is, the farther the communication distance, resulting in lower data rates at the same transmission power level and consequently lower energy efficiency. Therefore, we consider the condition where the above inequality becomes the equality $\sum_{i=1}^{N_{\mathrm{STR}}} \psi_i = \Theta$, and all relays should be positioned on the two shortest inferior arcs in this case \cite{wang2022stochastic}. These two arcs represent the shortest path along the Earth's surface from the ground transmitter $x_t$ to the ground receiver $x_r$ and the shortest path along the surface of the satellite constellation, from the first relay satellite $y_{m_1}$ to the last relay satellite $y_{m_{N_{\mathrm{STR}} - 1}}$. The latter's projection on the ground should overlap with the former. Since the azimuth angles of $x_t$ and $x_r$ are both $0$, all azimuth angles of the ideal relay positions are $0$. Note that satellites are generally not available precisely when the azimuth angles are $0$, hence these relay positions are considered ideal.

\par
The next step involves proving that the central angles in odd hops are equal, that is, $\psi_1 = \psi_3 = \dots = \psi_{N_{\mathrm{STR}} - 1}$. First of all, assume that $\psi_1 + \psi_3 + \dots + \psi_{N_{\mathrm{STR}} - 1} = \Psi \leq \Phi$. Then, we split $\mathscr{P}_{\mathrm{STR}}$ into two sub-problems to separately maximize the energy efficiency for odd hops and even hops. The sub-problem corresponding to odd hops can be formulated in (\ref{opt-A}) at the top of the next page,
\begin{table*}
\hrule
\begin{subequations}\label{opt-A}
	\begin{alignat}{2}
		\underset{\psi_1, \psi_3, \dots, \psi_{N_{\mathrm{STR}} - 1}}{\mathrm{minimize}} \ \ \ &  \sum_{i=1,i {\mathrm{\, is \, odd}}}^{ N_{\mathrm{STR}}-1 } \frac{B_{\rm{ST}}}{\rho_t^{(1)}} \log_2 \left( 1 + \frac{1}{\sigma_g^2} \rho_r^{(1)} \left( \sqrt{R_{\oplus}^2 + R_s^2 - 2 R_{\oplus} R_s \cos\psi_i } \right) \right) \label{opt2-1} \\
		\mathrm{subject \ to} \ \ \ \  & \ \ \ \ \ \ \ \ \ \ \ \ \ \ \ \ \ \ \ \ \ \ \ \ \sum_{i=1,i {\mathrm{\, is \, odd} }}^{N_{\mathrm{STR}}-1} \psi_i = \Psi, \label{opt2-2}
	\end{alignat}
\end{subequations}
\hrule
\end{table*}
where $\sqrt{R_{\oplus}^2 + R_s^2 - 2 R_{\oplus} R_s \cos\psi_i} = l_i$ represents the communication distance of the $i^{th}$ hop. Furthermore, the Lagrange multiplier $\mu$ is introduced to solve this constrained problem and let the equation (\ref{AppA-3}) at the top of the next page be satisfied.
\begin{equation}\label{AppA-3}
\begin{split}
    & \frac{\partial}{\partial \psi_i} \Bigg( \sum_{i=1,i {\mathrm{\, is \, odd}}}^{ N_{\mathrm{STR}}-1 } \frac{B_{\rm{ST}}}{\rho_t^{(1)}}  \\
    & \times \log_2 \bigg( 1 + \frac{1}{\sigma_g^2} \rho_r^{(1)} \left( \sqrt{R_{\oplus}^2 + R_s^2 - 2 R_{\oplus} R_s \cos\psi_i } \right) \bigg) \\  
    & \ \ \ \ \ \ \ \ \ \ \ \ \ \ \ \ \ \ \ \ \ \ \ \ \ - \mu \left( \sum_{i=1,i {\mathrm{\, is \, odd}}}^{ N_{\mathrm{STR}}-1 } \psi_i - \Psi \right) \Bigg) = 0. 
\end{split}
\end{equation}
The following differential equation exists for $\forall \, 1 \leq i \leq N_{\mathrm{STR}}-1$, and $i$ is odd:
\begin{sequation}
    \frac{\partial}{\partial \psi_i}  \frac{B_{\rm{ST}}}{\rho_t^{(1)}} \log_2 \left( 1 + \frac{1}{\sigma_g^2} \rho_r^{(1)} \left( \sqrt{R_{\oplus}^2 + R_s^2 - 2 R_{\oplus} R_s \cos\psi_i } \right) \right) = \mu. 
\end{sequation}
Therefore, there exists a $\mu$ satisfies the above equation set when $\psi_1 = \psi_3 = \dots = \psi_{N_{\mathrm{STR}} - 1}$. At this moment, denote $\psi_i$ with the equal value as $\theta_{{\mathrm{ISR}},1}$. Note that the above derivation process applies to any value taken by $\Psi$. Likewise, for the optimization of even hops, we can obtain $\psi_2 = \psi_4 = \dots = \psi_{N_{\mathrm{STR}}} = \theta_{{\mathrm{ISR}},2}$ through the same procedures.

\section{Proof of Lemma~\ref{lemma1}} \label{app:lemma1}
In the proof, we first consider the scenario of signal transmission from the GW to the satellite. Based on the relationship between the Euclidean distance of a hop and its central angle $l = \sqrt{R_{\oplus}^2 + R_s^2 - 2 R_{\oplus} R_s \cos\theta}$ provided in (\ref{opt2-1}), the average energy efficiency can be written as
\begin{sequation}\label{AppB-1}
\begin{split}
    &\overline{E}_{\mathrm{hop}}^{(1)} (\theta) = \mathbbm{E}_{W_{\mathrm{ST}}} \left[ E_{{\mathrm{ISR}},1} \right] \\
    & \overset{(a)}{\gtrsim} \frac{1}{\rho_t^{(1)}} B_{\rm{ST}} \log_2 \left( 1 +  \frac{\rho_t^{(1)} \zeta_{\rm{ST}} G_{\rm{ST}} \mathbbm{E}[W_{\mathrm{ST}}] }{ \left( R_{\oplus}^2 + R_s^2 - 2R_{\oplus}R_s \cos\theta \right) \sigma_s^2 } \left( \frac{\lambda_{\rm{ST}}}{4\pi} \right)^2 \right) \\
    & \overset{(b)}{=} \frac{1}{\rho_t^{(1)}} B_{\rm{ST}} \log_2 \left( 1 +  \frac{\rho_t^{(1)} \zeta_{\rm{ST}} G_{\rm{ST}} (2b_0 + \Omega)}{ \left( R_{\oplus}^2 + R_s^2 - 2R_{\oplus}R_s \cos\theta \right) \sigma_s^2 } \left( \frac{\lambda_{\rm{ST}}}{4\pi} \right)^2 \right),
\end{split}
\end{sequation}
where step $(a)$ follows Jensen's inequality given that $E_{{\mathrm{ISR}},1}$ is a convex function of $W_{\mathrm{ST}}$, and we obtain approximate results for the case when the equality holds. Step $(b)$ holds because the expectation of SR fading $\mathbbm{E}[W_{\mathrm{ST}}] = 2b + \Omega$, which can refer to the first-order moment of power gain for the SR fading \cite{abdi2003new}. The proof steps of the case where the signal is transmitted by a satellite and received by a GW are similar to the procedure in (\ref{AppB-1}), therefore omitted here.

\par
The derivation of the average energy efficiency in ISL differs from the above process in only two aspects. The first difference lies in the relationship between ISL distances and their corresponding central angles: $l = 2R_s \sin(\theta/2)$. The second difference concerns the expectation of the small-scale fading,
\begin{equation}
\begin{split}
    & \mathbb{E}[W_{\mathrm{SS}}] = \int_0^{A_0} \int_0^{\infty} w f_{W_{\mathrm{SS}}|\theta_d}\left ( w \right ) f_{\theta_d}\left ( \theta_d \right ) \mathrm{d} \theta_d \mathrm{d} w \\
    & = \int_0^{A_0}  \frac{\eta_s^2 w^{\eta_s^2 } }{A_0^{\eta_s^2}} \mathrm{d} w \int_0^{\infty} \cos\theta_d \frac{\theta_d}{\varsigma^2}\exp\left ( -\frac{\theta_d^2}{2\varsigma^2} \right ) \mathrm{d} \theta_d \\
    & \overset{(c)}{\approx} \frac{A_0 \eta_s^2}{1 + \eta_s^2} \left( 1 - \frac{1}{2} \int_0^{\infty} \frac{\theta_d^3}{\varsigma^2}\exp\left ( -\frac{\theta_d^2}{2\varsigma^2} \right ) \mathrm{d}\theta_d \right) \\ 
    & \overset{(d)}{=} \frac{A_0 \eta_s^2}{1 + \eta_s^2} \left( 1 - \int_0^{\infty} z\varsigma^2  \exp\left ( -z \right ) \mathrm{d}z \right) \overset{(e)}{=} \frac{A_0 \eta_s^2}{1 + \eta_s^2} \left( 1 - \varsigma^2 \right).
\end{split}
\end{equation}
Since $\theta_d$ is generally a small value, step $(c)$ follows the second-order Taylor expansion of $\cos\theta_d\approx 1 - \frac{\theta_d^2}{2}$. Step $(d)$ is derived by the substitution of $z = {\theta_d^2}/{2 \varsigma^2}$. As for step $(e)$, we take the expectation of the exponential distribution.

\section{Proof of Lemma~\ref{lemma2}} \label{app:lemma2}
In this appendix, we begin deriving the analytical expression of the distance scaling factor from the simplest scenario, that is, the first and last hops of STR and ISR. According to Slivnyak's theorem \cite{feller1991introduction}, the distribution of a homogeneous BPP is invariant with the rotation. To simplify the derivation, we assume that the ground transmitter or receiver is located at $\left( R_{\oplus},\theta,0 \right)$ and the ideal relay position of the satellite is located at $\left( R_s,0,0 \right)$. Therefore, the central angle of the first or last hop is $\theta$.

\par
Now, we need to determine the distributions of the azimuth angle and polar angle of the satellite closest to the ideal relay position. It is easy to notice that the satellite's azimuth angle $\varphi$ is uniformly distributed between $0$ and $2\pi$. Denote $\mathcal{S}(\psi)$ as the spherical cap with a central angle as $2\psi$, whose rotation axis is the line connecting the center of the Earth and the relay position at $\left( R_s,0,0 \right)$. The CDF of the satellite's polar angle $\psi_1$ is given by
\begin{equation}\label{AppC-1}
\begin{split}
    & F_{\psi_1}(\psi) = \mathbb{P}\left[ \psi_1 \leq \psi \right]  = 1 - \mathbb{P}\left[ {\mathcal{N}\left( {{\mathcal{S}(\psi)}} \right) = 0} \right] \\
    & = 1 - \left( 1 - \frac{\mathcal{A}\left( \mathcal{S}(\psi) \right)}{\mathcal{A}\left( \mathcal{S}(\pi) \right)} \right)^{N_s} = 1 - \left( \frac{ 1 + \cos\psi }{2} \right)^{N_s},
%= 1 - \left( 1 - \frac{ 2\pi R_s^2 (1-\cos\psi)}{4\pi R_s^2} \right)^{N_s}
\end{split}
\end{equation}
where $\mathcal{N}\left( \mathcal{S}(\psi) \right)$ counts the number of satellites in the spherical cap $\mathcal{S}(\psi)$, and $\mathcal{A}\left( \mathcal{S}(\psi) \right)$ denotes the area of $\mathcal{S}(\psi)$. Furthermore, the PDF of $\psi_1$ is
\begin{equation}\label{AppC-2}
\begin{split}
    f_{\psi_1}(\psi) = \frac{\mathrm{d}}{\mathrm{d}\psi} F_{\psi_1}(\psi) = \frac{N_s \sin\psi}{2} \left( \frac{ 1 + \cos\psi }{2} \right)^{N_s-1}.
\end{split}
\end{equation}
Next, the average distance from $\left( R_{\oplus}, \theta, 0 \right)$ to the relay satellite can be obtained by traversing the location of the nearest satellite,
\begin{equation}\label{AppC-3}
    \overline{d}^{(1)} (\theta) = \int_0^{2\pi} \int_0^{\pi} \frac{1}{2\pi} f_{\psi_1}(\psi) \, d \left( R_s,\psi,\varphi; R_{\oplus},\theta,0 \right) \mathrm{d}\psi \mathrm{d}\varphi,
\end{equation}
where $d \left( R_s,\psi,\varphi; R_{\oplus},\theta,0 \right)$ is the Euclidean distance between position $(R_s,\psi,\varphi)$ and position $(R_{\oplus},\theta,0)$, which is provided in (\ref{d_12}). From the definition, the distance scaling factor of the first and last hop in STR and ISR is given by,
\begin{equation}\label{AppC-4}
\begin{split}
    & \alpha^{(1)} \left(\theta\right) = \frac{ \overline{d}^{(1)} (\theta) }{ \sqrt{R_{\oplus}^2 + R_s^2 - 2 R_{\oplus} R_s \cos\theta} }.
    %= \frac{N_s}{4 \pi \sqrt{R_{\oplus}^2 + R_s^2 - 2 R_{\oplus} R_s \cos\theta}} \\
    %& \times \int_0^{2\pi} \int_0^{\pi} \sin\psi \left( \frac{ 1 + \cos\psi }{2} \right)^{N_s-1} d \left( R_s,\psi,\varphi; R_{\oplus},\theta,0 \right) \mathrm{d}\psi \mathrm{d}\varphi.
\end{split}
\end{equation}

\par
Then, we extend the above results to the distance scaling factor of middle hops in STR. In this case, both the relay satellite and relay GW corresponding to this hop deviate from the ideal relay positions. We first consider that the position of one of the relays is fixed. Without loss of generality, we assume the satellite's position is fixed at $\left( R_s, \theta, 0 \right)$, and the nearest GW to the ideal relay position at $\left( R_{\oplus}, 0 ,0 \right)$ is selected. Following the above steps, we can similarly obtain the distance scaling factor for this particular situation, 
\begin{equation}\label{AppC-5}
\begin{split}
    & \widetilde{\alpha}^{(1)} \left(\theta\right) = \frac{N_g}{4 \pi \sqrt{R_{\oplus}^2 + R_s^2 - 2 R_{\oplus} R_s \cos\theta}} \int_0^{2\pi} \int_0^{\pi} \sin\psi \\
    & \times \left( \frac{ 1 + \cos\psi }{2} \right)^{N_s-1} d \left( R_{\oplus},\psi,\varphi; R_s,\theta,0 \right) \mathrm{d}\psi \mathrm{d}\varphi.
\end{split}
\end{equation}
To derive an approximate distance scaling factor, we assume that the increments generated are independent and sequential. Increments firstly occur due to the deviation from the GW, followed by increments due to the satellites. Therefore, $d \left( R_{\oplus},\psi,\varphi; R_s,\theta,0 \right)$ in (\ref{AppC-5}) is substituted by $\alpha^{(1)}(\theta) \, d \left( R_{\oplus},\psi,\varphi; R_s,\theta,0 \right)$ after the first round's increment is applied, and the derivation of the second type of distance scaling factor is finished.

\par
The last is the distance scaling factor for middle hops in ISR. We similarly fix one relay satellite at the ideal relay position and consider the increments resulting from the random distribution of another relay satellite. Since the increments caused by both relay satellites are consistent, the final result is the square of the single satellite's increment. The proof process is similar to that of the middle hops of STR, therefore omitted here.

\section{Proof of Theorem~\ref{theorem1}}\label{app:theorem1}
To apply single-hop availability probability ion deriving routing availability probability, we need to determine the values of three variables $(\theta_1, \theta_2, \theta_c)$. The first two central angles are easily solvable, while $\theta_c$ is a random variable, leading us to categorize and discuss it separately. 
\begin{itemize}
\item First and last hops for STR: As $\theta_c$ is independent of the satellite relays' positions, we assume the first and last satellite relays are exactly located at the ideal relay positions. Therefore, $\theta_c = \theta_1 + \widetilde{\alpha}_g (\theta_2) \, \theta_2$ for the first hop, and $\theta_c = \theta_2 + \widetilde{\alpha}_g (\theta_1) \, \theta_1$ for the last hop. $\widetilde{\alpha}_g(\theta)$ is the scale factor that presents the increase in the central angle due to deviations of GW relay positions from the ideal relay positions, which is given in (\ref{theo1-2}).
\item Middle hops for STR: When $\theta_c$ presents the central angles between two GWs, $\theta_c = \widetilde{\alpha}_g (\theta_1) \, \theta_1 + \widetilde{\alpha}_g (\theta_2) \, \theta_2$. Otherwise, when $\theta_c$ presents the central angles between two satellites, $\theta_c = \widetilde{\alpha}_s (\theta_1) \, \theta_1 + \widetilde{\alpha}_s (\theta_2) \, \theta_2$, where $\widetilde{\alpha}_s(\theta)$ is the scale factor that presents the increase in the central angle due to deviations of satellites, which is also given in (\ref{theo1-2}).
\end{itemize}
The $\theta_c$ in ISR can be derived through a similar process and, therefore omitted here. Due to the independent distribution of points in BPP, we approximate that the availability probability for each hop is also independent of others. Hence, the routing availability probability is the product of the available probabilities for each hop. So far, Theorem~\ref{theorem1} has been proved.

\section{Proof of Lemma~\ref{lemma4}} \label{app:lemma4}
This proof starts with deriving the central angle distributions of the first hop and last hops in STR and ISR. Similar to Appendix~\ref{app:lemma2}, we rotate the coordinate system such that the position of the ground transmitter is $\left( R_{\oplus}, \phi, 0 \right)$, and the ideal position of the first relay satellite is $\left( R_s, 0, 0 \right)$. By traversing the potential positions of the nearest satellite, the CDF of central angle distribution can be derived as
\begin{sequation}
\begin{split}
    & F_{\theta_c^{(1)}} \left( \theta_c \, | \, \phi \right) = \int_0^{2\pi} \int_0^{\theta_c} R_s^2 \sin\psi \\ 
    & \times f_{\theta_1, \varphi_1}\left( \arccos \left( \frac{ R_s^2 + R_{\oplus}^2 -  ( d( R_s,\psi,\varphi; R_{\oplus},\phi,0))^2 }{2 R_s R_{\oplus}} \right) \right)  \mathrm{d}\psi \mathrm{d}\varphi,
\end{split}
\end{sequation}
where $d( R_s,\psi,\varphi; R_{\oplus},\theta_{{\mathrm{STR}},1},0)$ is defined in (\ref{d_12}). The probability that the nearest satellite's polar angle $\theta_1$ and azimuth angle $\varphi_1$ satisfies $\psi \leq \theta_1 < \psi + \mathrm{d}\psi$ and $\varphi \leq \varphi_1 < \psi + \mathrm{d}\varphi$ is
\begin{equation}
    f_{\theta_1, \varphi_1}(\theta) = \frac{f_{\psi_1}(\theta)}{2\pi R_s^2 \sin{\theta}} = \frac{N_s}{4\pi R_s^2} \left( \frac{ 1 + \cos\theta }{2} \right)^{N_s-1},
\end{equation}
where ${f_{\psi_1}(\theta)}$ is defined in (\ref{AppC-2}). As for the central angle distribution of middle hops in STR, $F_{\theta_c^{(2)}} \left( \theta_c \, | \, \phi \right)$ can be derived as,
\begin{equation}\label{AppE-3}
\begin{split}
    & F_{\theta_c^{(2)}} \left( \theta_c \, | \, \phi \right) = \mathbb{P}\left[ \cos\theta_c^{(2)} > \cos\theta_c \, | \, \phi \right] \\
    & = \mathbb{P}\left[ \left( \alpha^{(1)}(\phi) \, d^{(1)} \right)^2 > R_s^2 + R_{\oplus}^2 - 2R_s R_{\oplus} \cos\theta_c \right] \\
    & = \mathbb{P}\bigg[ R_s^2 + R_{\oplus}^2 - 2R_s R_{\oplus} \cos\theta_c^{(1)} > \\
    & \left( \alpha^{(1)}(\phi) \right)^{-2} \left( R_s^2 + R_{\oplus}^2 - 2R_s R_{\oplus} \cos\theta_c \right) \bigg] \\
    & = F_{\theta_c^{(1)}} \bigg( \arccos \bigg( \frac{1}{2R_s R_{\oplus}} \bigg( R_s^2 + R_{\oplus}^2 - \left( \alpha^{(1)}(\phi) \right)^{-2} \\
    & \times \left( R_s^2 + R_{\oplus}^2 - 2R_s R_{\oplus} \cos\theta_c \right) \bigg) \bigg) \, \bigg| \, \phi \bigg).
\end{split}
\end{equation}
The proof of middle hops for ISR is similar to that of STR, therefore omitted here. The PDF of the contact angle distribution can be derived by can be obtained by taking the derivative of the CDF with respect to $\theta_c$.

\section{Proof of Theorem~\ref{theorem3}}\label{app:theorem3}
According to Definition~\ref{defenergy}, the routing energy efficiency can be obtained by taking the expectation of the random variables. The energy efficiency for the $i^{th}$ hop, where $2 \leq i \leq N_{\mathrm{ISR}}$ (a middle hop), is given as,
\begin{equation}\label{AppF-1}
\begin{split}
    & E_{{\mathrm{ISR}},3}(\theta_{{\mathrm{ISR}},3}) = \mathbb{E}_{\theta_i} \bigg[ \mathbb{E}_{W_{\rm{SS}}} \bigg[ \frac{1}{\beta \, \rho_t^{(3)}} B_{\rm{SS}} \\
    & \times \log_2 \left( 1 + \frac{\rho_r^{(3)} ( 2R_s \sin(\theta_i / 2) )}{\sigma_s^2} \right) \bigg] \bigg] 
    \\
    & = \int_0^{\pi} \int_0^{A_0} \int_0^{\infty} \frac{B_{\rm{SS}}}{\beta \, \rho_t^{(3)}} f_{\theta_c^{(3)}}(\psi \, | \,\theta_{{\mathrm{ISR}},3}) f_{W_{\rm{SS}}|\theta_d}(w) \\
    & \times f_{\theta_d}(\theta_d) \log_2 \left( 1 + \frac{\rho_r^{(3)} ( 2R_s \sin(\psi/2) )}{\sigma_s^2} \right) \mathrm{d}\theta_d \, \mathrm{d}w \, \mathrm{d}\psi.
\end{split}
\end{equation}
From this, it can be seen the derivation of the accurate energy efficiency expression involves a quadruple integral ($f_{\theta_c^{(3)}}(\psi \, | \,\theta_{{\mathrm{ISR}},3})$ contains a single integral). This introduces considerable computational complexity, thus we substitute the portion involving the expectation of small-scale fading with the average energy efficiency provided in Lemma~\ref{lemma1}. The energy efficiency of a middle hop can be approximated as
\begin{equation}
\begin{split}
    \widetilde{E}_{{\mathrm{ISR}},i}(\theta_{{\mathrm{ISR}},3}) =  \int_0^{\pi} f_{\theta_c^{(3)}}(\psi \, | \,\theta_{{\mathrm{ISR}},3}) \overline{E}_{\mathrm{hop}}^{(3)}(\theta_{{\mathrm{ISR}},3}) \mathrm{d}\psi.
\end{split}
\end{equation}
The same derivation approach can also be used for the first and last hops. The specific derivation process and expressions are ignored here.

\bibliographystyle{IEEEtran}
\bibliography{references}

\begin{thebibliography}{10}
\providecommand{\url}[1]{#1}
\csname url@samestyle\endcsname
\providecommand{\newblock}{\relax}
\providecommand{\bibinfo}[2]{#2}
\providecommand{\BIBentrySTDinterwordspacing}{\spaceskip=0pt\relax}
\providecommand{\BIBentryALTinterwordstretchfactor}{4}
\providecommand{\BIBentryALTinterwordspacing}{\spaceskip=\fontdimen2\font plus
\BIBentryALTinterwordstretchfactor\fontdimen3\font minus
  \fontdimen4\font\relax}
\providecommand{\BIBforeignlanguage}[2]{{%
\expandafter\ifx\csname l@#1\endcsname\relax
\typeout{** WARNING: IEEEtran.bst: No hyphenation pattern has been}%
\typeout{** loaded for the language `#1'. Using the pattern for}%
\typeout{** the default language instead.}%
\else
\language=\csname l@#1\endcsname
\fi
#2}}
\providecommand{\BIBdecl}{\relax}
\BIBdecl

\bibitem{yue2022security}
P.~Yue, J.~An, J.~Zhang, G.~Pan, S.~Wang, P.~Xiao, and L.~Hanzo, ``On the
  security of {LEO} satellite communication systems: Vulnerabilities,
  countermeasures, and future trends,'' \emph{\rm{available online:}
  https://arxiv.org/abs/2201.03063}.

\bibitem{chaudhry2020free}
A.~U. Chaudhry and H.~Yanikomeroglu, ``Free space optics for next-generation
  satellite networks,'' \emph{IEEE Consumer Electronics Magazine}, 2020.

\bibitem{zhu2021integrated}
X.~Zhu and C.~Jiang, ``Integrated satellite-terrestrial networks toward {6G}:
  Architectures, applications, and challenges,'' \emph{IEEE Internet of Things
  Journal}, vol.~9, no.~1, pp. 437--461, 2021.

\bibitem{chaudhry2022temporary}
A.~U. Chaudhry and H.~Yanikomeroglu, ``Temporary laser inter-satellite links in
  free-space optical satellite networks,'' \emph{IEEE Open Journal of the
  Communications Society}, vol.~3, pp. 1413--1427, 2022.

\bibitem{radhakrishnan2016survey}
R.~Radhakrishnan, W.~W. Edmonson, F.~Afghah, R.~M. Rodriguez-Osorio, F.~Pinto,
  and S.~C. Burleigh, ``Survey of inter-satellite communication for small
  satellite systems: Physical layer to network layer view,'' \emph{IEEE
  Communications Surveys \& Tutorials}, vol.~18, no.~4, pp. 2442--2473, 2016.

\bibitem{del2019technical}
I.~Del~Portillo, B.~G. Cameron, and E.~F. Crawley, ``A technical comparison of
  three low earth orbit satellite constellation systems to provide global
  broadband,'' \emph{Acta astronautica}, vol. 159, pp. 123--135, 2019.

\bibitem{wang2022ultra}
R.~Wang, M.~A. Kishk, and M.-S. Alouini, ``Ultra-dense {LEO} satellite-based
  communication systems: {A} novel modeling technique,'' \emph{Communications
  Magazine}, vol.~60, no.~4, pp. 25--31, 2022.

\bibitem{9348676_1}
J.~Hu, L.~Cai, C.~Zhao, and J.~Pan, ``Directed percolation routing for
  ultra-reliable and low-latency services in low earth orbit ({LEO}) satellite
  networks,'' in \emph{IEEE 92nd Vehicular Technology Conference
  (VTC2020-Fall)}, 2020, pp. 1--6.

\bibitem{shen2020dynamic_2}
L.~Shen, Y.~Wang, L.~Liu, S.~Liu, D.~Wang, Y.~Fan, H.~Zhou, and T.~Ling, ``A
  dynamic modified routing strategy based on load balancing in {LEO} satellite
  network,'' in \emph{International Conference on Wireless and Satellite
  Systems}.\hskip 1em plus 0.5em minus 0.4em\relax Springer, 2020, pp.
  233--244.

\bibitem{geng2021agent_3}
S.~Geng, S.~Liu, Z.~Fang, and S.~Gao, ``An agent-based clustering framework for
  reliable satellite networks,'' \emph{Reliability Engineering \& System
  Safety}, vol. 212, p. 107630, 2021.

\bibitem{knight2011internet}
S.~Knight, H.~X. Nguyen, N.~Falkner, R.~Bowden, and M.~Roughan, ``The internet
  topology zoo,'' \emph{IEEE Journal on Selected Areas in Communications},
  vol.~29, no.~9, pp. 1765--1775, 2011.

\bibitem{yang2016towards}
Y.~Yang, M.~Xu, D.~Wang, and Y.~Wang, ``Towards energy-efficient routing in
  satellite networks,'' \emph{IEEE Journal on Selected Areas in
  Communications}, vol.~34, no.~12, pp. 3869--3886, 2016.

\bibitem{rabjerg2021exploiting}
J.~W. Rabjerg, I.~Leyva-Mayorga, B.~Soret, and P.~Popovski, ``Exploiting
  topology awareness for routing in {LEO} satellite constellations,'' in
  \emph{Global Communications Conference (GLOBECOM)}.\hskip 1em plus 0.5em
  minus 0.4em\relax IEEE, 2021, pp. 1--6.

\bibitem{zhao2021multi_4}
N.~Zhao, X.~Long, and J.~Wang, ``A multi-constraint optimal routing algorithm
  in {LEO} satellite networks,'' \emph{Wireless Networks}, pp. 1--12, 2021.

\bibitem{8068282_5}
Y.~Cao, Y.~Shi, J.~Liu, and N.~Kato, ``Optimal satellite gateway placement in
  space-ground integrated network for latency minimization with reliability
  guarantee,'' \emph{IEEE Wireless Communications Letters}, vol.~7, no.~2, pp.
  174--177, 2018.

\bibitem{Al-1}
A.~Al-Hourani, ``An analytic approach for modeling the coverage performance of
  dense satellite networks,'' \emph{IEEE Wireless Communications Letters},
  vol.~10, no.~4, pp. 897--901, 2021.

\bibitem{ok-2}
N.~Okati and T.~Riihonen, ``Modeling and analysis of {LEO} mega-constellations
  as nonhomogeneous {P}oisson point processes,'' in \emph{IEEE 93rd Vehicular
  Technology Conference (VTC2021-Spring)}, 2021, pp. 1--5.

\bibitem{yastrebova2020theoretical}
A.~Yastrebova, I.~Angervuori, N.~Okati, M.~Vehkaper{\"a}, M.~H{\"o}yhty{\"a},
  R.~Wichman, and T.~Riihonen, ``Theoretical and simulation-based analysis of
  terrestrial interference to {LEO} satellite uplinks,'' in \emph{IEEE Global
  Communications Conference (GLOBECOM)}.\hskip 1em plus 0.5em minus 0.4em\relax
  IEEE, 2020, pp. 1--6.

\bibitem{ok-1}
N.~Okati, T.~Riihonen, D.~Korpi, I.~Angervuori, and R.~Wichman, ``Downlink
  coverage and rate analysis of low {E}arth orbit satellite constellations
  using stochastic geometry,'' \emph{IEEE Transactions on Communications},
  vol.~68, no.~8, pp. 5120--5134, 2020.

\bibitem{talgat2020stochastic}
A.~Talgat, M.~A. Kishk, and M.-S. Alouini, ``Stochastic geometry-based analysis
  of {LEO} satellite communication systems,'' \emph{IEEE Communications
  Letters}, vol.~25, no.~8, pp. 2458--2462, 2021.

\bibitem{al-3}
A.~Al-Hourani, ``A tractable approach for predicting pass duration in dense
  satellite networks,'' \emph{IEEE Communications Letters}, vol.~25, no.~8, pp.
  2698--2702, 2021.

\bibitem{al2021session}
------, ``Session duration between handovers in dense {LEO} satellite
  networks,'' \emph{IEEE Wireless Communications Letters}, vol.~10, no.~12, pp.
  2810--2814, 2021.

\bibitem{242206}
C.-S. Choi, ``Modeling and analysis of downlink communications in a
  heterogeneous {LEO} satellite network,'' \emph{IEEE Transactions on Wireless
  Communications}, 2024, early Access.

\bibitem{wang2022stochastic}
R.~Wang, M.~A. Kishk, and M.-S. Alouini, ``Stochastic geometry-based low
  latency routing in massive {LEO} satellite networks,'' \emph{IEEE
  Transactions on Aerospace and Electronic Systems}, vol.~58, no.~5, pp.
  3881--3894, 2022.

\bibitem{wang2023reliability}
------, ``Reliability analysis of multi-hop routing in multi-tier {LEO}
  satellite networks,'' \emph{IEEE Transactions on Wireless Communications},
  2023, early Access.

\bibitem{feller1991introduction}
W.~Feller, \emph{An introduction to probability theory and its applications,
  Volume 2}.\hskip 1em plus 0.5em minus 0.4em\relax John Wiley \& Sons, 1991,
  vol.~81.

\bibitem{gopal2014modulation}
P.~Gopal, V.~Jain, and S.~Kar, ``Modulation techniques used in
  {E}arth-to-satellite and inter-satellite free space optical links,'' in
  \emph{Unmanned/Unattended Sensors and Sensor Networks X}, vol. 9248.\hskip
  1em plus 0.5em minus 0.4em\relax SPIE, 2014, pp. 210--219.

\bibitem{ata2022performance}
Y.~Ata and M.-S. Alouini, ``Performance of integrated ground-air-space {FSO}
  links over various turbulent environments,'' \emph{IEEE Photonics Journal},
  vol.~14, no.~6, pp. 1--16, 2022.

\bibitem{wang2022conditional}
R.~Wang, A.~Talgat, M.~A. Kishk, and M.-S. Alouini, ``Conditional contact angle
  distribution in {LEO} satellite-relayed transmission,'' \emph{IEEE
  Communications Letters}, vol.~26, no.~11, pp. 2735--2739, 2022.

\bibitem{sag2018modelling}
E.~Sag and A.~Kavas, ``Modelling and performance analysis of 2.5 gbps
  inter-satellite optical wireless communication (isowc) system in {LEO}
  constellation.'' \emph{J. Commun.}, vol.~13, no.~10, pp. 553--558, 2018.

\bibitem{lou2023coverage}
Z.~Lou, B.~E.~Y. Belmekki, and M.-S. Alouini, ``Coverage analysis of hybrid
  {RF/THz} networks with best relay selection,'' \emph{IEEE Communications
  Letters}, vol.~27, no.~6, pp. 1634--1638, 2023.

\bibitem{lin2022refracting}
Z.~Lin, H.~Niu, K.~An, Y.~Wang, G.~Zheng, S.~Chatzinotas, and Y.~Hu,
  ``Refracting {RIS}-aided hybrid satellite-terrestrial relay networks: Joint
  beamforming design and optimization,'' \emph{IEEE Transactions on Aerospace
  and Electronic Systems}, vol.~58, no.~4, pp. 3717--3724, 2022.

\bibitem{he2022multiobjective}
Y.~He, Y.~Liu, C.~Jiang, and X.~Zhong, ``Multiobjective anti-collision for
  massive access ranging in {MF-TDMA} satellite communication system,''
  \emph{IEEE Internet of Things Journal}, vol.~9, no.~16, pp. 14\,655--14\,666,
  2022.

\bibitem{abdi2003new}
A.~Abdi, W.~C. Lau, M.-S. Alouini, and M.~Kaveh, ``A new simple model for land
  mobile satellite channels: {F}irst-and second-order statistics,'' \emph{IEEE
  Transactions on Wireless Communications}, vol.~2, no.~3, pp. 519--528, 2003.

\end{thebibliography}

\end{document}